%% file: main.tex
\documentclass[5p,times]{elsarticle}

\usepackage{amssymb}
\usepackage{amsthm}
\usepackage{todonotes}
\usepackage{doi}
\usepackage{graphicx}
\usepackage{amsmath}
\usepackage{physics}
\usepackage{upgreek}
\usepackage{subcaption}
\usepackage{comment}
\usepackage{listings}
\usepackage{soul}
\usepackage{xcolor}
\usepackage[linguistics]{forest}
\usepackage{letltxmacro}
\usepackage{tikz}
\usetikzlibrary{quantikz}
\usepackage{multirow}
\usepackage[acronym,toc,shortcuts]{glossaries}
\usepackage{booktabs} 
\usepackage{svg}
\usepackage{adjustbox}

\usepackage{tablefootnote}

\usepackage{pifont}% http://ctan.org/pkg/pifont

\definecolor{ashgrey}{rgb}{0.7, 0.75, 0.71}

\usetikzlibrary{shapes.geometric}
\tikzset{
    gateTriangle/.style={
        isosceles triangle,
        draw,
        fill=white,
        inner sep=0.03cm
    },
    gateTriangleInv/.style={
        isosceles triangle,
        rotate=-180,
        draw,
        fill=white,
        inner sep=0.05cm
    }
}

\DeclareExpandableDocumentCommand{\gateTriangle}{O{} m}{|[gateTriangle,#1]| #2 \qw}
\DeclareExpandableDocumentCommand{\gateTriangleInv}{O{} m}{|[gateTriangleInv,#1]| #2 \qw}

\newcommand*{\SavedLstInline}{}
\LetLtxMacro\SavedLstInline\lstinline
\DeclareRobustCommand*{\lstinline}{%
  \ifmmode
    \let\SavedBGroup\bgroup
    \def\bgroup{%
      \let\bgroup\SavedBGroup
      \hbox\bgroup
    }%
  \fi
  \SavedLstInline
}

\definecolor{codegreen}{rgb}{0,0.6,0}
\definecolor{codegray}{rgb}{0.5,0.5,0.5}
\definecolor{codepurple}{rgb}{0.58,0,0.82}
\definecolor{backcolour}{rgb}{0.95,0.95,0.92}

\lstdefinestyle{mystyle}{
    backgroundcolor=\color{backcolour},   
    commentstyle=\color{codegreen},
    keywordstyle=\color{magenta},
    numberstyle=\tiny\color{codegray},
    stringstyle=\color{codepurple},
    basicstyle=\ttfamily\footnotesize,
    breakatwhitespace=false,         
    breaklines=true,                 
    captionpos=b,                    
    keepspaces=true,                 
    numbers=left,                    
    numbersep=5pt,                  
    showspaces=false,                
    showstringspaces=false,
    showtabs=false,                  
    tabsize=2
}

\input{acronyms}

\usepackage{titlesec}

\titleformat{\paragraph}
{\normalfont\normalsize\itshape}{\theparagraph}{0.2em}{\normalfont\normalsize\itshape}
\titlespacing{\paragraph}
{0pt}{*1}{*1}[0.2pt]
\begin{document}

\begin{frontmatter}

\title{Review of Distributed Quantum Computing. From single QPU to High Performance Quantum Computing}

\author[cesga]{David Barral}
\author[citius]{F. Javier Cardama}
\author[cesga]{Guillermo Díaz}
\author[cesga]{Daniel Faílde}
\author[cesga]{Iago F. Llovo}
\author[cesga]{Mariamo Mussa Juane}
\author[citius]{Jorge Vázquez-Pérez}
\author[cesga]{Juan Villasuso}

\author[dec]{César Piñeiro}
\author[cesga]{Natalia Costas}
\author[citius,dec]{Juan C. Pichel}
\author[citius,dec]{Tomás F. Pena}
\author[cesga]{Andrés Gómez}

\affiliation[cesga]{organization={Galicia Supercomputing Center (CESGA)},
            addressline={Avda.\ de Vigo S/N}, 
            city={Santiago de Compostela},
            postcode={15705}, 
            %state={A Coruña},
            country={Spain}}
\affiliation[citius]{organization={Centro Singular de Investigación en Tecnoloxías Intelixentes (CiTIUS)},
            addressline={Universidade de Santiago de Compostela}, 
            city={Santiago de Compostela},
            postcode={15782}, 
            %state={A Coruña},
            country={Spain}}
\affiliation[dec]{organization={Departamento de Electrónica e Computación},
            addressline={Universidade de Santiago de Compostela}, 
            city={Santiago de Compostela},
            postcode={15782}, 
            %state={A Coruña},
            country={Spain}}
\begin{abstract}

The emerging field of quantum computing has shown it might change how we process information by using the unique principles of quantum mechanics. As researchers continue to push the boundaries of quantum technologies to unprecedented levels, distributed quantum computing raises as an obvious path to explore with the aim of boosting the computational power of current quantum systems. This paper presents a comprehensive survey of the current state of the art in the distributed quantum computing field, exploring its foundational principles, landscape of achievements, challenges, and promising directions for further research. From quantum communication protocols to entanglement-based distributed algorithms, each aspect contributes to the mosaic of distributed quantum computing, making it an attractive approach to address the limitations of classical computing. Our objective is to provide an exhaustive overview for experienced researchers and field newcomers. 
\end{abstract}


\end{frontmatter}

\input{Section1}

\input{Section2}

\input{Section3}

\input{Section4}

\input{Section5}

\input{Conclusions}

\bibliographystyle{elsarticle-num}

\bibliography{checked-references}

\end{document}

%% file: acronyms.tex
\newacronym{dqc}{DQC}{distributed quantum computing}
\newacronym{qpu}{QPU}{Quantum Processing Unit}
\newacronym{pu}{PU}{Processing Unit}
\newacronym{hpc}{HPC}{High-Perfomance Computing}

\newacronym{nv}{NV}{Nitrogen Vacancy}

\newacronym{epr}{EPR}{Einstein, Podolski and Rosen}
\newacronym{dv}{DV}{discrete variable}
\newacronym{cv}{CV}{continuos variable}
\newacronym{bsm}{BSM}{Bell-state measurement}
\newacronym{eit}{EIT}{electro\-magnetically-induced transparency}
\newacronym{nisq}{NISQ}{Noisy Intermediate-Scale Quantum}
\newacronym{qkd}{QKD}{Quantum Key Distribution}
\newacronym{spdc}{SPDC}{spontaneous parametric down-conversion}
\newacronym{crib}{CRIB}{controlled reversible inhomogeneous broadening}
\newacronym{afc}{AFC}{atomic frequency combs}
\newacronym{edfa}{EDFA}{erbium-doped fiber amplifiers}
\newacronym{eaqn}{EAQN}{En\-tan\-gle\-ment-assisted Quantum Network}
\newacronym{qlan}{QLAN}{Quantum Local Area Network}
\newacronym{qm}{QM}{quantum memory}
\newacronym{qn}{QN}{quantum network}
\newacronym{ip}{IP}{Internet Protocol}
\newacronym{qwan}{QWAN}{Quantum Wide Area Network}
\newacronym{egp}{EGP}{Entanglement Generation Protocol}
\newacronym{mhp}{MHP}{Midpoint Heralding Protocol}
\newacronym{qmm}{QMM}{Quantum Memory Management}
\newacronym{ghz}{GHZ}{Green\-ber\-ger-Hor\-ne-Zei\-lin\-ger}

\newacronym{osi}{OSI}{Open Systems Interconnection}
\newacronym{qi}{QI}{Quantum Internet}
\newacronym{qkdn}{QKDN}{Quantum Key Distribution Network}
\newacronym{qin}{QIN}{Quantum Information Network}
\newacronym{pmn}{PMN}{Prepare and Measurement Network}
\newacronym{edn}{EDN}{Entanglement Distribution Network}
\newacronym{qmn}{QMN}{Quantum Memory Network}
\newacronym{fqn}{FQN}{Fault Tolerant Qubit Network}
\newacronym{irtf}{IRTF}{Internet Research Task Force}
\newacronym{qrna}{QRNA}{Quantum Recursive Network Architecture}

\newacronym{ir}{IR}{Intermediate Representation}
\newacronym{isa}{ISA}{Instruction Set Architecture}
\newacronym{qram}{qRAM}{quantum Random Access Memory}
\newacronym{vqe}{VQE}{Variational Quantum Eigensolver}
\newacronym{qir}{QIR}{Quantum Intermediate Representation}
\newacronym{qIR}{qIR}{quantum Intermediate Representation}
\newacronym{mlir}{MLIR}{Multi-Level Intermediate Representation}
\newacronym{qiro}{QIRO}{Quantum Intermediate Representation for Optimization}
\newacronym{ssa}{SSA}{Static Single Assigment}
\newacronym{qssa}{QSSA}{Quantum Static Single Assigment}
\newacronym{sqir}{SQIR}{Small Quantum Intermediate Representation}
\newacronym{xacc}{XACC}{eXtreme-scale Accelerator programming framework}
\newacronym{qbir}{QBIR}{Quantum Block Intermediate Representation}
\newacronym{qisa}{QISA}{Quantum Instruction Set Architecture}
\newacronym{rpo}{RPO}{Relaxed Peephole Implementation}
\newacronym{ecc}{ECC}{Equivalent Circuit Classes}
\newacronym{voqc}{VOQC}{Verified Optimizer for Quantum Circuit}
\newacronym{inquir}{InQuIR}{Intermediate Representation for Interconnected Quantum Computers}
\newacronym{qpalg}{QPAlg}{Quantum Process Algebra}
\newacronym{eqpalg}{eQPAlg}{Extended Quantum Process Algebra}
\newacronym{mpi}{MPI}{Message Passing Interface}
\newacronym{qmpi}{QMPI}{Quantum MPI}
\newacronym{qft}{QFT}{Quantum Fourier Transform}
\newacronym{alu}{ALU}{arithmetic logic unit}
\newacronym{fpu}{FPU}{floating-point unit}
\newacronym{qasm}{QASM}{Quantum Assembly Language}
\newacronym{qnpu}{QNPU}{Quantum Network Processing Unit}
\newacronym{qa-dqc}{QA-DQC}{qubit allocation problem for distributed quantum computing}
\newacronym{mhsa}{MHSA}{multistage hybrid simulated annealing}
\newacronym{oee}{OEE}{Overall Extreme Exchange}
\newacronym{roee}{rOEE}{relaxed-OEE}
\newacronym{kahypar}{KaHyPar}{Karlsruhe Hypergraph Partitioning}
\newacronym{kl}{KL}{Kernighan-Lin}
\newacronym{hqa}{HQA}{Hungarian Qubit Assigment}
\newacronym{fgp}{FGP}{Fine Grained Partitioning}
\newacronym{drl}{DRL}{Deep Reinforment Learning}
\newacronym{ppo}{PPO}{Proximal Policy Optimization}
\newacronym{qubo}{QUBO}{Quadratic Unconstrained Binary Optimization}
\newacronym{cdap}{CDAP}{Community Detection Assistant Partitioning}
\newacronym{frp}{FRP}{Fair and Reliable Partitioning}
\newacronym{dis}{DIS}{Delayed Instruction Scheduling}
\newacronym{amp}{AMP}{Adaptive Multi-Programming}
\newacronym{qdca}{QDCA}{Quantum Divide and Conquer Algorithm}
\newacronym{tdag}{TDAG}{Tree-based Directed Acyclic Graph}
\newacronym{gagdo}{GAGDO}{Genetic Algorithm for Global Gate Direction Optimization}
\newacronym{mis}{MIS}{Maximum Independent Set}
\newacronym{sdk}{SDK}{Software Development Kit}

\newacronym{vqa}{VQA}{Variational Quantum Algorithm}
\newacronym{qpe}{QPE}{Quantum Phase Estimation}
\newacronym{qml}{QML}{Quantum Machine Learning}
\newacronym{qaoa}{QAOA}{Quantum Approximate Optimization Algorithm}
\newacronym{qps}{QPS}{quasi-probabilistic simulation}
\newacronym{qpd}{QPD}{quasi-probabilistic decomposition}
\newacronym{locc}{LOCC}{local operations and classical communication}
\newacronym{nme}{NME}{non-maximally-entangled}
\newacronym{tn}{TN}{Tensor Networks}

%% file: Section1.tex
\section{Introduction}
\label{sec:introduction}
In the pursuit of achieving superior computational abilities, quantum computing has arisen as a promising frontier with huge potential. While individual quantum systems have shown impressive capabilities, the idea of distributed quantum computing introduces a new approach that could vastly increase computational power. 
This study aims to explore in depth the current landscape of \gls*{dqc}, also known in certain literature as modular quantum computation, from physical devices and interconnection networks to distributed algorithms. In this review, we will analyze the different solutions proposed and the challenges posed by this rapidly advancing field.

As we examine distributed quantum systems more closely, it becomes clear that collaborative and interconnected quantum processors are essential for overcoming the constraints faced by standalone systems. Problems of both fundamental origin --~decoherence, dissipation, and crosstalk~-- and practical origin --~processor topology, cabling, connectors, and control electronics~-- hinder the fabrication of ultra-large \glspl*{qpu}~\cite{Vanmeter2008}. It is thus foreseeable in the short term that quantum computers will not scale in a local device with a large number of qubits in a single quantum processor. A distributed infrastructure with several quantum processors that contain a limited number of qubits could overcome this difficulty. In fact, there is almost a consensus among both the academic community and companies that the practical realization of large-scale quantum processors should adopt a distributed approach based on clusters of small, modular quantum chips within a network infrastructure, with classical and/or quantum communications \cite{Quera2023, Quantumnews2024, Ionq2024}. 
\glspl*{qpu} are intended to be seamlessly integrated into a classical \gls*{hpc} infrastructure, alongside CPUs, GPUs, and other hardware accelerators~\cite{Saurabh2023AMiddleware,wintersperger2022qpu,vazquez2024qpu, McCaskey2019XACC:Computing, mccaskey2021-qcor}. This integration allows for their utilization in collaboration within a shared development environment, leading to what is already called quantum-centric supercomputing centers~\cite{Gambetta.2022}.

As an example of this trend, IBM recently unveiled Quantum System Two~\cite{Gambetta.2023}, a modular architecture that will serve as the basis for building their new quantum-centric \gls*{hpc} infrastructures. The model unveiled features three IBM Quantum Heron processors, each with 133 fixed-frequency qubits and tunable couplers. According to IBM, Heron yields a 3-5x improvement in performance with respect to the previous 127-qubit Eagle processor, virtually eliminating crosstalk.

However, the interest in \gls*{dqc} is not new. We have to go back to the end of the 20th century to find the first works that analyzed the possibility of using non-local effects to perform distributed computing~\cite{grover1997quantum,cleve1997substituting}.
This interest grew after Cirac et al.'s work, where it was shown that \gls*{dqc} is superior to classical computing for the phase estimation problem even under non-ideal conditions~\cite{Cirac1999}. Shortly after, Eisert et al.\ \cite{eisert2000optimal} and Collins et al.\ \cite{Collins2001} took a step forward introducing resource-opti\-mi\-zed protocols for non-local quantum gates, necessary to move from specific problems like phase estimation to universal quantum computing. At the same time, DiVincenzo~\cite{Divicenzo2000} included, in his famous criteria for a quantum computer, two additional no-so-well-known items related to \gls*{dqc} and the interconnection of \glspl*{qpu}: the ability to interconnect stationary and ﬂying qubits, and to faithfully transmit ﬂying qubits between speciﬁed locations.

After the first theoretical studies on the feasibility of \gls*{dqc}, a series of proposals for experimental realizations began to appear gradually~\cite{Lim2005,Serafini2006,Jiang2007,Oi2006}. At the same time, several interesting developments regarding \gls*{dqc} algorithms were made, such as the distributed versions of the Grover and Shor algorithms~\cite{Gupta2007, Yimsiriwattana2004Shor}. 
The first taxonomy of \gls*{dqc} systems was proposed by Yepez~\cite{Yepez2001type} in the early 2000s, where two types of systems were described: those with entanglement between nodes, called type-I, and those with only inter-node classical communication, called type-II. Jozsa and Linden later demonstrated that a type-II quantum computer cannot achieve exponential speedup when the computation requires entanglement across the full set of qubits~\cite{Jozsa2003}.

\begin{figure}[t]
    \centering
    \includegraphics[width=0.8\linewidth]{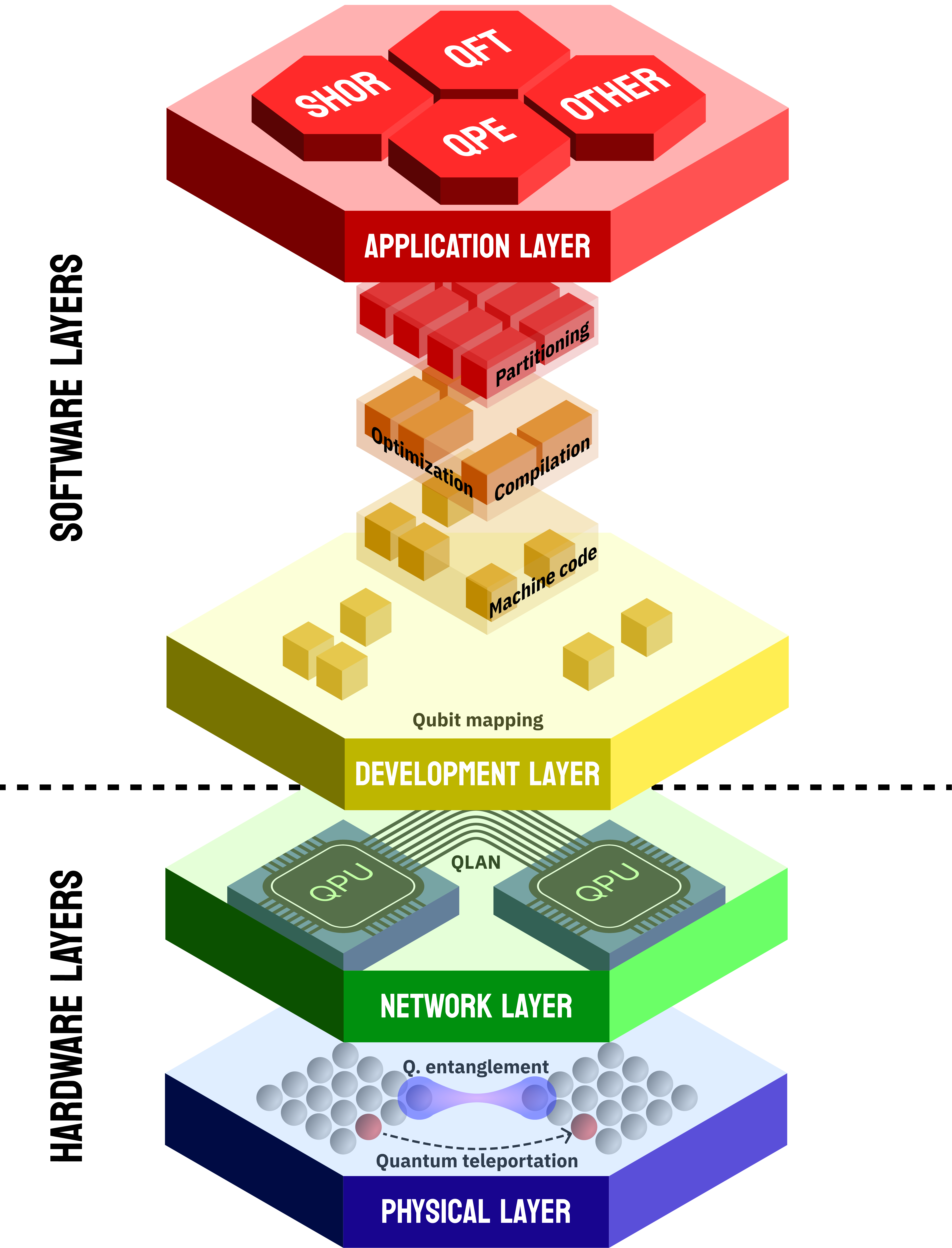}
    \caption{Layered model for distributed quantum computing.}
    \label{fig:intro:layers}
\end{figure}

Considering these initial works as a starting point, this review extensively examines the current advancements in the field of \gls*{dqc}, extending and updating previous surveys on this subject \cite{Caleffi2022DistributedSurvey,Cacciapuoti2020QuantumComputing}. This review provides an in-depth analysis of the latest proposals in the field of DQC, including all the full-stack, from the communications level to distributed applications. It investigates the fundamental principles, accomplishments, challenges, and potential directions for future exploration. 

To facilitate the readers' understanding, this survey is structured according to a layered model, as depicted in Figure \ref{fig:intro:layers}, similar to the full-stack architecture presented by \cite{Rodrigo2021OnComputers} or the abstract model in \cite{cuomo2020towards}. 

The two lower layers Fig.\ \ref{fig:intro:layers} encompass the hardware developments needed to implement a distributed quantum system and would be equivalent to the three lower layers of the classical OSI model. So, the physical layer refers to the mechanisms that allow two physically separated \glspl*{qpu} to be connected, while the network layer defines how to establish communication between multiple \glspl*{qpu}. Directly above this layer, we discuss advances in development tools that allow applications to be distributed and executed on a distributed quantum system, including partitioning, compilation, optimization, and mapping algorithms.
Finally, in the uppermost layer, we address distributed algorithms. It is important to note that these layers are interdependent, with each layer influencing those immediately preceding and succeeding it. For instance, the development of a compiler is influenced by the underlying hardware and also provides support for different partitioning techniques in the application layer. 

Following this structure, the review is organized as follows. Section \ref{sec:physical} describes the available quantum mechanical tools to transmit quantum information. We then present in Section \ref{sec:networking} proposals oriented to the creation of networks interconnecting multiple \glspl*{qpu}. Next, Section \ref{compiler:sec:distributed-quantum-compilers} discusses solutions that allow applications to run in distributed environments, including partitioning, distribution, compilation, and mapping techniques. Section \ref{sec:Aplications} presents different proposals for applications running in these environments. We will end the paper with a summary of the current state of the art and open lines in the field.

%% file: Section2.tex
\section{Physical layer for distributed quantum computing}
\label{sec:physical}

\gls*{dqc} aims at performing arbitrary computational tasks between unknown quantum states at the distant nodes of a quantum network. These networks, identically to their classical counterparts, coordinate and distribute information across devices. However, quantum networks have multiple features and limitations that make these tasks difficult, primarily arising from the {\it no-cloning} theorem: arbitrary quantum states cannot be \textit{perfectly} copied; therefore, quantum information cannot be replicated and broadcast \cite{Wootters1982}. Fortunately, the properties of quantum systems can be exploited in a way that allows us to circumvent this impediment and reliably transmit quantum information or control quantum systems remotely. This section will briefly describe which quantum mechanical tools are available for this purpose.

First and foremost, the physical resource that enables performing non-local computation is \textit{entanglement}, a unique correlation of joint quantum systems stronger than any classical counterpart but very fragile, hard to create and to maintain long.
Entanglement lies at the heart of quantum communications, facilitating the distribution of quantum states encoding quantum information through a protocol known as \textit{quantum teleportation} or \textit{teledata}.
Multiple teleportation variants exist, which are designed to either transmit data in one direction --~\textit{quantum teleportation} or \textit{teledata}~-- but also bi-directional communication --~\textit{entanglement swapping}~-- and gate operation at a distance --~\textit{gate teleportation} or \textit{telegate}. Furthermore, the basic two-node teleportation can be extended to multi-party distribution networks composed of a large number of nodes. Some parties may either help the rest of the network in the quantum communication protocol --~\textit{assisted teleportation}~--, or the quantum information may be imperfectly broadcast from one sender to the rest --~\textit{quantum telecloning}.

In the following sections, we will introduce these protocols in detail.

\subsection{Quantum entanglement}

A system of two spatially separated quantum particles with maximally correlated momenta and maximally anti-correlated positions --~dubbed EPR pair~-- is the basis of the thought experiment on the nonlocality of quantum mechanics proposed in 1935 by \gls*{epr} \cite{Einstein1935}. This challenging idea led to the birth of the concept of quantum entanglement \cite{Schrodinger1935} which is now recognized as one of the three primary forms of quantum correlations: entanglement \cite{Werner1989}, steering \cite{Wiseman2007} and Bell non-locality \cite{Bell1964}. 

Entanglement is the property of a quantum system that illustrates the impossibility of describing a composed system in terms of just its individual components due to nonclassical correlations of certain degree(s) of freedom of the subsystems \cite{Horodecki2009}. Typical examples of these degrees of freedom are the position and momentum of free particles, the polarization of light, energy levels of trapped ions, or transverse atomic spins. These degrees of freedom are related to observables that present a discrete and finite spectrum or a continuous and infinite one. Hence, the terms \gls*{dv} and \gls*{cv}. This review focuses on \gls*{dv} because it is the most common in quantum computing. 

Archetypical examples of \gls*{dv} entangled quantum states are the pure states 
\begin{equation} \label{bell}
    \begin{split}
    \vert\Phi^{\pm}\rangle &=\frac{1}{\sqrt{2}}\left(\vert 0\rangle_A \vert 0\rangle_B \pm \vert 1\rangle_A \vert 1 \rangle_{B}\right), \\
    \vert\Psi^{\pm}\rangle &=\frac{1}{\sqrt{2}}\left(\vert 0\rangle_A \vert 1\rangle_B \pm \vert 1\rangle_A \vert 0 \rangle_{B}\right),
    \end{split}
\end{equation}
dubbed Bell states, where two parties --~Alice and Bob--~ share two qubits A and B encoded in a dichotomic degree of freedom as polarization, spin, or any other two-level quantum variable \cite{Gerry2005}. A perfect non-local correlation arises as Alice's measurement outcome determines Bob's measurement outcome. This property allows us to build an intuition of how Bell states are a natural choice for quantum communication: if a quantum gate, whose matrix representation is symmetric, is applied to one of the qubits of the Bell state $\vert\Phi^{+}\rangle$, it is the same as if the gate was applied to the other qubit. The gate somewhat ‘slides’ between qubits through the entanglement, like beads on a string \cite{Andres2018}.

These entangled states are the basis of a large number of quantum information protocols, one of which is quantum teleportation, which we introduce in the following section. 

\subsection{Quantum teleportation or teledata}
\label{ch6:quantum-teleportation}

Teleportation is a popular concept in pop culture and has been featured in countless books, movies, TV shows, and video games. It is the process of instantaneously moving an object or person from one location to another, typically without traversing the space in between. Thirty years ago, a quantum information protocol based on a similar concept --~dubbed quantum teleportation~-- was introduced in a landmark paper \cite{Bennett1993}. This quantum protocol enables the reconstruction of an unknown quantum state of a given physical system at a different location without actually transmitting the system. Quantum teleportation requires two key ingredients: 
\begin{itemize}
    \item \textit{Quantum entanglement}, the essential resource without which it would be impossible within the constraints of quantum mechanics.
    \item \textit{Classical communication between the locations}, which excludes superluminal communication. 
\end{itemize}

Quantum teleportation plays a pivotal role in the development of quantum technologies \cite{Acin2018}. It overcomes some of the limitations of quantum communications and quantum computing using the non-local transfer of unknown information. Quantum teleportation networks \cite{Karlsson1998}, entanglement swapping \cite{Zukowski1993}, and quantum repeaters \cite{Briegel1998} enable the distribution of entanglement over long distances \cite{Ren2017}, while quantum gate teleportation \cite{Gottesman1999} and measurement-based quantum computing \cite{Raussendorf2001} are examples of techniques that distribute local gate operations among physically disconnected parties \cite{Chou2018}. 

Proof-of-principle demonstrations of quantum teleportation were successfully achieved using diverse physical substrates as photonic qubits \cite{Bouw1997}, optical modes \cite{Furusawa1998}, atomic ensembles \cite{Sherson2006}, nuclear magnetic resonance \cite{Nielsen1998}, trapped atoms \cite{Riebe2004, Barrett2004}, and solid-state systems \cite{Steffen2013}. Over the last years, the focus has moved to teleporting more complex states --~larger number of degrees of freedoms or higher dimension qubits \cite{Wang2015, Hu2020}~-- and to real-world applications in quantum communications and computation \cite{Ren2017, Llewellyn2019, Hoke2023}.

\begin{figure}[t]
    \centering
    \begin{subfigure}{\linewidth}
        \centering
        \includegraphics[width=0.8\linewidth]{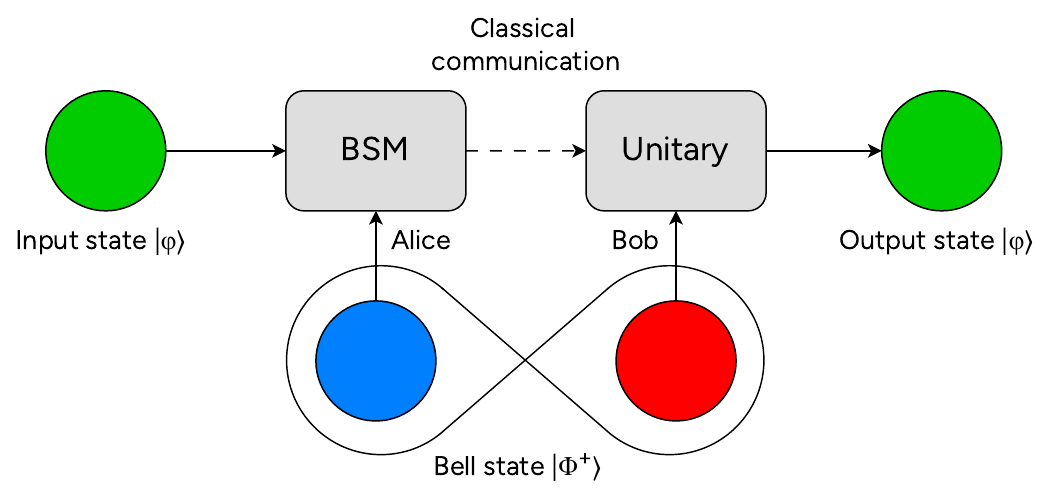}
        \caption{Quantum teleportation.}
        \label{subfigure:chapter3:quantum_teleportation}
    \end{subfigure}
    \hfill
    \begin{subfigure}{\linewidth}
        \centering
        \includegraphics[width=0.8\linewidth]{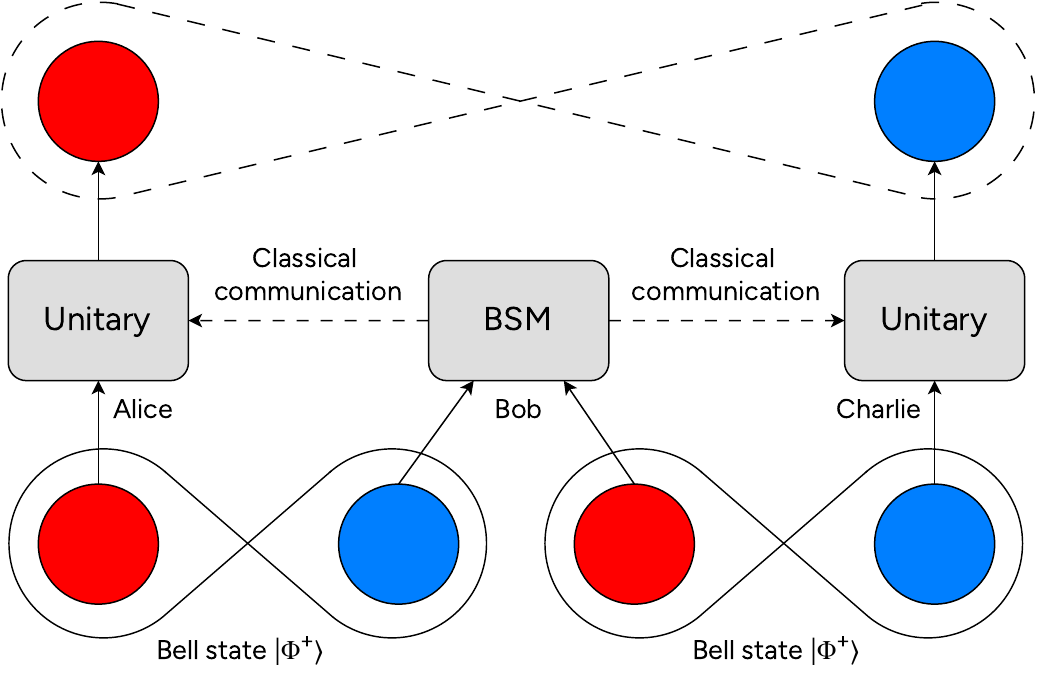}
        \caption{Entanglement swapping.}
        \label{subfigure:chapter3:entanglement_swapping}
    \end{subfigure}
    \hfill
    \begin{subfigure}{\linewidth}
        \centering
        \includegraphics[width=0.8\linewidth]{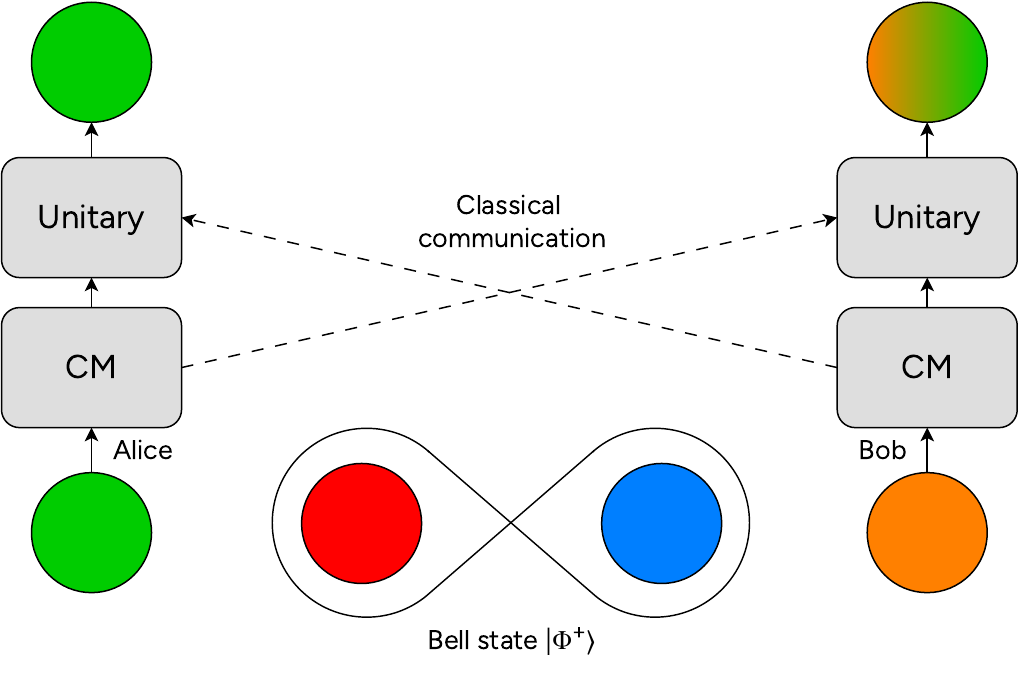}
        \caption{Gate teleportation.}
        \label{subfigure:chapter3:gate_teleportation}
    \end{subfigure}
            
    \caption{Sketch of quantum communication protocols: (a) Quantum-state teleportation (teledata), (b) entanglement swapping, and (c) quantum-gate teleportation (telegate). BSM: Bell-state measurement. CM: controlled operation and projective measurement.}
    \label{fig:quantum_communication}
\end{figure}

\begin{figure*}[t]
    \centering
    \begin{subfigure}[b]{0.45\linewidth}
        \centering
        \begin{adjustbox}{scale=0.7}
        \begin{quantikz}
            \lstick[2]{$\text{QPU}_1$} && \lstick{$|a\rangle$}& \ctrl{1}\gategroup[wires=3, steps=4, style={dashed, rounded corners, fill=green!10, inner sep=2pt}, background]{Teleport}                                                                                           & \gate{H} & \meter{}          & \cwbend{2} & \\
                                       && \lstick[2]{$|\Phi^+\rangle$}   & \targ{}  & \qw      & \meter{}          &            & \\
            \lstick[4]{$\text{QPU}_2$} &&                                & \qw      & \qw      & \gate{X} \vcw{-1} & \gate{Z}   & \swap{1} & \qw        & \qw        & \qw \\
                                       && \lstick{$|0\rangle$}           & \qw      & \qw      & \qw               & \qw        & \targX{} & \ctrl{1} \gategroup[wires=3, steps=2, style={dashed, rounded corners, fill=red!10, inner sep=2pt}, background, label style={label position=below,anchor=north,yshift=-0.2cm}]{Local CZs}   & \ctrl{2}   & \qw \\
                                       && \lstick{$|t_1\rangle$}         & \qw      & \qw      & \qw               & \qw        & \qw      & \control{} & \qw        & \qw \\
                                       && \lstick{$|t_2\rangle$}         & \qw      & \qw      & \qw               & \qw        & \qw      & \qw        & \control{} & \qw 
        \end{quantikz}
        \end{adjustbox}
        \caption{Teledata.}
        \label{subfig:circuits_tele:teledata}
    \end{subfigure}

    \begin{subfigure}[b]{0.45\linewidth}
        \centering
         \begin{adjustbox}{scale=0.7}
         \begin{quantikz}
            \lstick[2]{$\text{QPU}_1$} && \lstick{$|a\rangle$}           & \ctrl{1}\gategroup[wires=3, steps=2, style={dashed, rounded corners, fill=green!10, inner sep=2pt}, background]{Cat-Ent}                                                                                & \qw               & \qw        & \qw        & \qw \gategroup[wires=3, steps=2, style={dashed, rounded corners, fill=yellow!10, inner sep=2pt}, background]{Cat-DisEnt}     & \gate{Z} \vcw{2} & \qw \\
                                       && \lstick[2]{$|\Phi^+\rangle$}   & \targ{}  & \meter{}          &            & \\
            \lstick[4]{$\text{QPU}_2$} &&                                & \qw      & \gate{X} \vcw{-1} & \ctrl{1} \gategroup[wires=3, steps=2, style={dashed, rounded corners, fill=red!10, inner                                                                                sep=2pt}, background,label style={label position=below,anchor=north,yshift=-0.2cm}]{Local CZs}   & \ctrl{2}   & \gate{H} & \meter{}         & \qw \\
                                       && \lstick{$|t_1\rangle$}           & \qw      & \qw               & \control{} & \qw        & \qw      & \qw              & \qw \\
                                       && \lstick{$|t_2\rangle$}         & \qw      & \qw      & \qw                 & \control{} & \qw      & \qw              & \qw 
        \end{quantikz}
        \end{adjustbox}
        \caption{Telegate.}
         \label{subfig:circuits_tele:telegate}
    \end{subfigure}
    \caption{Examples of teledata and telegate circuits for the application of CZs gates over $|t_1\rangle$ and $|t_2\rangle$ with the remote state $|a\rangle$ as control. (a) The state $|a\rangle$ in QPU\textsubscript{1} is teleported to the first qubit of QPU\textsubscript{2} (b) Cat-entangler and cat-disentangler primitives \cite{Yimsiriwattana2004Generalized} are used to implement the remote control.}
    \label{fig:circuits_teledata_telegate}
\end{figure*}
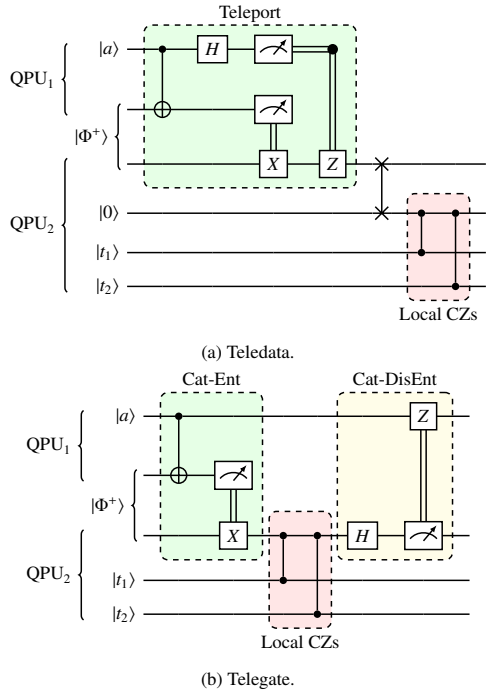

In the teledata protocol, Alice and Bob share an entangled Bell state as that given by Eq.\ \eqref{bell} \cite{Bouw1997}, see Figs.\ \ref{subfigure:chapter3:quantum_teleportation} and \ref{subfig:circuits_tele:teledata} in physical and circuit representations, respectively. A third party, commonly named Charlie, provides Alice with a qubit C to be teleported to Bob. Importantly, Charlie's qubit state $\rho$ is unknown to both Alice and Bob unlike in remote state preparation \cite{Bennett2001}. She then performs a \gls*{bsm}, which randomly projects with equal probability her qubits A and C into one of the four Bell states $\vert\Phi^{\pm}\rangle$ or $\vert\Psi^{\pm}\rangle$. As a result, Bob's qubit B is simultaneously projected onto the state $T^{\dag} \rho T$, where $T\in\{I, X, Z, Z X\}$ is an elementary or a combination of Pauli operators. As the last step, Alice informs Bob of the \gls*{bsm} outcome through the classical channel using two classical bits --~feed-forward~-- and Bob applies the suitable gate $T$ to his qubit to recover Charlie's unknown state $\rho$ at his location.

Regarding the figures of merit of quantum teleportation, there are mainly two: 
\begin{enumerate}
    \item The \textit{\gls*{bsm} efficiency} or Alice's success probability for distinguishing a complete basis of entangled states --~like the four Bell states. This varies for different information encodings: for instance, for a simple realization of Bell-state measurement using \gls*{dv} photonic qubits, the Bell efficiency is 50\% at maximum \cite{Weinfurter1994}.
    \item The \textit{teleportation fidelity} $F \in [0,1]$ between Charlie's input state and Bob's output state averaged over all Alice's measurement results and Charlie's input states. The benchmark for the teleportation fidelity is surpassing the fidelity for state transfer without quantum resources, using for instance just classical correlations, i.e., $F>F_{\text{class}}$, where $F_{\text{class}}=2/3$ for \gls*{dv} \cite{Massar1995}.
\end{enumerate}

Table \ref{Table2} shows examples of recent milestones in quantum teleportation in different technologies. More details on the state of the art can be found in \cite{Pirandola2015, Hu2023}. 

Quantum teleportation has seamlessly made the leap from laboratory conditions to real-world implementation in urban environments, showcasing its adaptability and robust functionality. Teleportation networks allow for the reliable transfer of quantum information between a number of distant nodes, even in the presence of non-ideal features as noise and loss. Recent advances include demonstrations of two-node teleportation over a metropolitan network \cite{Sun2016, Valivarthi2016}, links between nanophotonic memories and ion traps in an urban network \cite{Knaut2023, Krutyanskiy2023}, and multinode entanglement over a metropolitan network with a cloud of Rubidium atoms in a ring cavity acting as a quantum memory \cite{Liu2023}. More on quantum networks will be delved in Section \ref{sec:networking}.

\begin{table}[t]
\centering
\footnotesize
\renewcommand{\arraystretch}{1.25}
\begin{tabular}{l c c c c}
\hline\hline
\textbf{Quantum technol.} & \textbf{Bell eff.} & \textbf{Fidel.} & \textbf{Max. dist.} & \textbf{Memory} \\%[0.5ex]
\hline \hline
Polarization \cite{Ren2017} & 25\% & 0.80 & 1400 km & NA  \\
Integrated opt. \cite{Llewellyn2019} &   25\% &   0.894 &   10 m &   NA  \\
Superconduct. \cite{Chou2018} &   100\% & 0.79 & chip & 1 ms \\
Cavity QED \cite{Daiss2021, Langenfeld2021} &   100\%  & 0.833 & 60 m &   -- \\
Ion Trap \cite{Wan2019} &   100\% & 0.845 & chip &   -- \\
Rare-earth \cite{Lago-Rivera2021} &   50\% & 0.86 & 1 km & 17.5 µs \\
\hline
\end{tabular}
\caption{Some milestones in quantum teleportation in terms of Bell efficiency, fidelity, distance of teleportation, and quantum memory. QED: quantum electrodynamics.}
\label{Table2}
\end{table}

\subsection{Variants of quantum teleportation}

Quantum teleportation is a primitive of quantum information science and has a number of variants essential for \gls*{dqc}. In the following we review the most important three: entanglement swapping, quantum gate teleportation --~telegate~-- and multipartite teleportation.

\subsubsection{Entanglement swapping}

Entanglement swapping is a variant of quantum teleportation that enables remote correlations by the transfer of quantum entanglement between distant end-users that do not directly share a quantum resource. In this case, Bob shares two entangled states, one with Alice and the other with Charlie, as shown in Figure~\ref{subfigure:chapter3:entanglement_swapping}. Bob acts as a relay between them, performing Bell measurements and broadcasting the outcomes by a classical channel to them, who apply the suitable gates to their qubits. As a result, Alice and Charlie now share an entangled state conditioned on the result of Bob's measurement \cite{Zukowski1993}. This protocol, together with entanglement distillation\footnote{Entanglement distillation, aka entanglement purification, involves converting $N$ copies of any entangled state $\rho$ into a certain quantity of nearly pure Bell pairs, solely through local operations and classical communication.} \cite{Bennett1996}, enables the distribution of entanglement over large distances, being the basis of quantum repeaters \cite{Briegel1998}. Related to entanglement swapping are fusion gates \cite{Browne2005, Bartolucci2023}, where projective measurements probabilistically \textit{fuse} small entangled states in order to produce large entangled states --~cluster states~-- useful for measurement-based quantum computing \cite{Raussendorf2001}. 

The first demonstration of entanglement swapping was carried out by Pan et al. using polarization-entangled photons \cite{Pan1998}. Swapping has been recently applied to connect two spatially-separated solid-state quantum memories by telecom links \cite{Lago-Rivera2021}, and to entangle non-neighboring \gls*{nv} qubits in a multinode teleportation network \cite{Hermans2022}.

\subsubsection{Quantum gate teleportation or telegate}

In gate-based quantum computing, a sequence of unitary operations (usually single- and two-qubit) are applied on a set of qubits. However, sometimes there is no direct interaction between qubits on which we want to apply a two-qubit gate \cite{Jiang2007}. Quantum gate teleportation, also known as telegate, reduces the topological requirements by substituting two-qubit gates with other cost-effec\-ti\-ve resources: auxiliary entangled states, local measurements, and single-qubit operations \cite{Gottesman1999}. Typically, Alice and Bob want to perform a non-local operation on unknown control and target states using a shared Bell state as a quantum channel. To this end, both perform locally controlled operations and projective measurements (CM) on their half Bell state and control/target states. After this step, partial quantum information is transferred between the two parties conditioned to the measurement outcomes. Cross communication of the results through a classical channel enables Alice and Bob to perform suitable corrections to the control and target states. This procedure results in a controlled gate operation on two non-interacting input states --~see Figures~\ref{subfigure:chapter3:gate_teleportation} and \ref{subfig:circuits_tele:telegate} for physical and circuit representations, respectively. The first experimental demonstration of quantum gate teleportation was a remote CNOT operation carried out through photon entanglement and linear optical manipulations \cite{Huang2004}. Recent advances in remote operations comprise superconducting qubits, trapped ions, and quantum electrodynamics cavity nodes \cite{Chou2018, Wan2019, Daiss2021}.

When applied to multipartite entangled states with a given topology, suitable measurement on a given network node teleport unitary-transformed-state to other nodes. This is the basis of measurement-based quantum computing \cite{Raussendorf2001}. 

\subsubsection{Multipartite teleportation}

Multipartite entangled states as the \gls*{ghz} state enable a natural extension of quantum teleportation to more than two parties \cite{Pan2000}. These $N$-party protocols for multipartite teleportation enable two variants: assisted and unassisted teleportation --~commonly referred to as quantum telecloning. In the first case, \textit{assisted teleportation}, Alice \textit{helps} the communication between Bob and Charlie by performing a tailored measurement and broadcasting the result to them, thus improving the entanglement between them \cite{Karlsson1998}. In the second case, \textit{quantum telecloning}, Charlie teleports to Alice and Bob simultaneously, hence with a teleportation fidelity, limited by the no-cloning theorem, given by $F=(MN+M+N)/(MN+2M)$, for $N$ senders and $M$ receivers of qubits \cite{Murao1999}. 

Examples of assisted teleportation are open-destination teleportation \cite{Zhao2004} and, more recently, shared-quantum-secret teleportation \cite{Lee2020}. Quantum telecloning was, in turn, demonstrated in \gls*{dv} by means of partial teleportation \cite{Zhao2005}. Cloning of entanglement \cite{Peng2020} and copy distribution \cite{Wang2021} are recent examples of this variant of teleportation.

\subsection{Quantum devices for entanglement distribution}
\label{ch6:quantum-devices}
As no clear winner to the race to general purpose \glspl*{qpu} has been established, diverse quantum computing platforms are currently under development. Each competing technology has shown different advantages and disadvantages, such as short gate operation in superconducting \glspl*{qpu}; long qubit coherence in \gls*{nv} color centers in diamond, nuclear spin or ionic/Rydberg atom qubits and qubit mobility and straight-forward long-distance distribution in photonic systems, such as C-band photons in fiber optics. Despite the current lead of superconducting qubit systems in the \gls*{nisq} era, it is likely that no single technology will cover every need of quantum computing, with the capability of homogenizing the quantum computing platforms.

For this reason, modular architectures featuring specialized, single\nobreakdash-purpose hardware are currently under development. The aim is to maximize performance and demonstrate quantum advantage for distributed, scalable quantum computing systems \cite{Awschalom2021}. The quantum devices that are part of this network can be categorized in one of the following categories: i) \glspl*{qpu}, the singular devices where qubit operations take place to perform a quantum algorithm; ii) \textit{quantum transducers}, which transform variations in a quantum property of a system into a transmittable signal, connecting qubits of different kinds, e.g., spin-photon, or of the same kind but at a different frequency, e.g., microwave-optical photon; iii) \textit{quantum memories}, which maintain a quantum state or quantum entanglement over a long period of time, e.g., in trapped ions; iv) \textit{quantum repeaters}, which allow entanglement operations at a distance to be reliable and perform deterministic teleportation protocols, and v) \textit{entanglement routers and switches}, which allow the teleportation protocols to be performed between arbitrary parts of the distributed system, enabling true any-to-any connectivity.

This section will describe the aforementioned devices in detail and discuss the current research advances in each technology.

\subsubsection{Quantum transducers}
\label{ch6:transducers}
The communication between {\it local} qubits of systems where the quantum operations take place (e.g., \glspl*{qpu}, memories or repeaters) requires the conversion, or {\it transduction}, of their states to a different system used for delivery of quantum states in the form of {\it flying} qubits, which have the requirements of being highly mobile and well coupled to the specific local platform. Multiple flying qubit systems have been proposed, such as short-distance electronic states in semiconductor devices \cite{Edlbauer2022}, direct delivery of nuclei with long-lived nuclear-spin qubit encoding \cite{Zhong2015} and, more commonly, single photons, given their naturally mobile nature and their low coupling with the environment. In classical communications, the high-rate transfer of current technologies is only possible due to the high bandwidth and low attenuation of fiber optics, enabling the underwater connection of continents at tens of thousands of kilometers \cite{Hurst2000}. The current state-of-the-art telecommunication systems also implement multiplexing, i.e., encoding information at multiple wavelengths through the same fiber \cite{Grobe2013}. 

For the same reasons, single photons are also the most natural information carrier choice for the distribution of quantum states at a distance, and extensive research has focused on the accurate manipulation of photonic states using linear and nonlinear optical devices \cite{Munn1993, Barbieri2005}. For many applications such as \gls*{qkd} \cite{Bennett2014}, single photons are commonly approximated by strongly attenuated coherent states to encode qubits, which has been used to demonstrate the transmission of quantum states for \gls*{qkd} at speeds exceeding 110 Mbps at short distance, or up to 55~dB attenuation at long distance, equivalent to over 200 km of standard telecom fiber connection~\cite{Li2023}. A recent review on the topic of single-photon generation can be found in~\cite{MohammadNejad2023}. High-fidelity (up to 90\%), heralded teleportation of quantum states without the need for preformed Bell pairs has also been demonstrated by Langenfeld et al., which could potentially enable deterministic, short-distance, and low-latency quantum teleportation~\cite{Langenfeld2021}. However, further research is required to bring this method's quantum efficiency and fidelity closer to the entanglement distillation protocols.

Nevertheless, in order to distribute entanglement by transmitting a local quantum state, the flying qubits must be well coupled to the particular local quantum system. The protocols that can fulfill this task are generally referred to as {\it pitch-and-catch} protocols, in which the flying qubit is coupled to a local quantum system, either by direct emission or by interaction with the system. Finding physical mechanisms that can perform quantum transduction, the conversion of local qubits into quantum signals, has become an area of significant scientific-technological interest. Several solid-state to infrared single photons transducer mechanisms have been found~\cite{Castelleto2023}, e.g., in quantum dots~\cite{Arakawa2020}, diamond color centers~\cite{Aharonovich2011, Castelletto12, Nguyen2019}, rare-earth doped crystals~\cite{ZhongGoldner2019, Lago-Rivera2023} and trapped ions~\cite{Bock2018}; on the other hand, other transducer mechanisms have been shown from physical qubits to microwave photons, such as spin-photon coupling in Si double quantum dot spin qubits~\cite{Samkharadze18}. Using pitch-and-catch protocols, successful one-to-one entanglement distribution between neighboring entanglement nodes has also been demonstrated, e.g., arbitrary phonon coupling between individual ions in an ion trap~\cite{Hucul2015}, optical coupling of ion- or Rydberg atom-chains in optical cavities~\cite{Ramette2022}, or deterministic transmission of excitations between superconducting \glspl*{qpu} using cryogenic microwave waveguides~\cite{Kurpiers2018, Magnard2020, Renger2023}, demonstrating entanglement at a distance.

However, the most promising way of generating deterministic entanglement between remote systems is via entanglement swapping. This primarily consists of generating entanglement between flying qubits (most frequently photons) and local qubits (i.e., trapped ions, neutral atoms, or NV centers), then performing \gls*{bsm} on the photons of each pair. Hence, their joint wavefunction collapses in the same non-separable state, and the matter systems become entangled. 

Multiple techniques can be utilized to achieve the initial photon-matter qubit entanglement. On the one hand, correlated photon sources such as \gls*{spdc} or quantum dots can be used. \gls*{spdc} sources consist of a non-linear crystal pumped by a strong laser beam generating pairs of maximally entangled photons with some probability, which can then be frequency-filtered and made to interact with the physical qubits. Hyperentanglement, where more than one degree of freedom can simultaneously be maximally entangled (e.g., polarization and direction of two photons) has been demonstrated using this type of sources~\cite{Barbieri2005,Zhao2023}. Quantum dot-based sources have very attractive properties for this purpose, such as being triggered on-demand and energy-tunable~\cite{Li2023,Zhang2017,Ou2022}, and reaching fidelities over 90\% \cite{Hopfmann2021,Aumann2022}.

Individual photon-matter qubit entangled pairs can also be generated in certain systems, to then entangle the remote matter qubits via \gls*{bsm}. To this purpose, heralded entanglement of photons emitted after de-excitation from prepared excited states has been shown in trapped-ion qubits~\cite{Hucul2015, Moehring2007, Maunz2007}, neutral atoms~\cite{Hofmann12} and diamond \gls*{nv}-center qubits~\cite{Bernien2013, Pfaff2014, Kalb2017, Humphreys2018}. After the subsequent \gls*{bsm}, fidelities to Bell states of up to 88\% at 230~m have been demonstrated in trapped-ions~\cite{Krutyanskiy23}, and deterministic qubit state transfer between different \gls*{nv}-center nodes has also been shown~\cite{Hermans2022}.

\begin{figure}[t]
    \centering
    \includegraphics[width=0.6\linewidth]{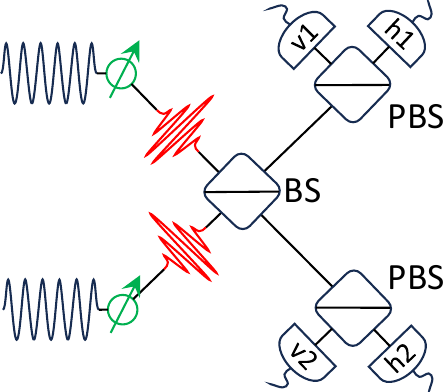}
    \caption{Diagram representing photonic entanglement swapping by Bell-state measurement. BS: beam splitter; PBS: polarizing beam splitter; h1, h2, v1, v2: single photon detectors.}
    \label{fig:bsm}
\end{figure}

Figure~\ref{fig:bsm} shows a schematic representation of light-matter entanglement swapping by \gls*{bsm}. Matter qubits (green) emit single photons (red), entangling one of their respective degrees of freedom (e.g., spin and polarization) with each other in a superposition of states. The successful projection onto a Bell-state is then heralded by detector coincidences, which results in the matter qubits becoming entangled. Clicks in v1 and h1 (or v2 and h2) herald the creation of a $\ket{\Psi^+}$ state, while clicks in v1 and v2 (or h1 and h2) herald a $\ket{\Psi^-}$ state.

Moreover, interconnecting quantum systems may require coupling platforms that operate at different photon frequencies. For this purpose, techniques are being developed to implement frequency conversion of single photons on demand, maintaining certain properties (such as polarization) intact, which would enable the transcoding of qubits between platforms. One such technique is heralded up-conversion from infrared to visible light, which has been achieved through sum frequency generation in nonlinear crystals~\cite{Zhou2016, Kaiser2019}. More recently, Murakami et al.~\cite{Murakami2023} have demonstrated frequency conversion from visible to infrared using pairs of non-degenerate photons generated by \gls*{spdc}, and Weaver et al.~\cite{Weaver2023} have shown frequency bidirectional transduction from microwave to infrared light using transduction assisted by a resonant mechanical mode. However, the quantum efficiency of these techniques is currently low and significant efforts are underway to push it towards unity. In addition to the aforementioned frequency conversion techniques, recent work by Sahu et al.~\cite{Sahu2023} has demonstrated deterministic entanglement between the quadratures of propagating microwave and optical photons in cryogenic waveguides, a first step towards interconnecting superconducting qubits with long-range communication systems and memories.

\subsubsection{Quantum memories}
\label{ch6:memories}
To fully take advantage of the entanglement distribution and distillation protocols for both short and long distance quantum communication, it is paramount that the coherence time of the communication qubits is longer than the protocol itself, surviving multiple rounds of qubit exchange and entanglement purification. These long-lived qubits, organized as large registries, are known as quantum memories or \glspl*{qram}. 

The simplest quantum memories are photonic memories, in which photons are stored and then retrieved after a given time. Multiple approaches exist, such as using free space optical loops triggered by heralding \cite{Pittman2002} or fiber delay lines~\cite{Landry2007} and cavities with tunable Q-factor~\cite{Leung2006, Tanabe2007}. Stimulated photon-echo is a more advanced technique based on the absorption and delayed reemission of single photons with the same quantum state after an ensemble of atoms is rephased~\cite{Moiseev2004, Lvovsky2009, Guo2023}, which has been demonstrated e.g., using slow light by \gls*{eit}~\cite{Gorshov2007}, \gls*{crib}~\cite{Tittel2010} and \gls*{afc} in rare-earth doped crystals~\cite{Azfelius2009,Ortu2022}. All-photonic systems (i.e., photonic quantum computing) can already take advantage of photonic memories, as they do not require transduction~\cite{Knill2001,Kok2007}. 

However, both the difficulty of retrieving single photons with high fidelity as well as the low scalability of photonic-based memories have pushed forward extensive research on multiple alternative quantum memory technologies, demonstrating high-fidelity single-qubit gates in excess of the threshold needed for quantum error correction~\cite{Knill1998,Kitaev2003}. Notable examples are trapped-ion and -neutral atom qubits, which use the hyperfine structure of atomic ensembles of ions~\cite{Harty2014}, or neutral alkali or alkaline earth single atoms in optical tweezers~\cite{Schlosser2001,Norcia2018,Barnes2022} to encode the quantum states, which can be individually addressed by microwave pulses~\cite{Isenhower2010}. Quantum memories based on diamond \gls*{nv}-centers have also been demonstrated (see~\cite{Doherty2013} and references therein). Some of these technologies have demonstrated long coherence times, of up to 10 minutes in single trapped-ion qubits~\cite{Wang2017} and up to six hours in cryogenically cooled Eu$^{3+}$-doped yttrium orthosilicate nuclear spin qubits~\cite{Zhong2015}. More recently, Barnes et al.\ \cite{Barnes2022} have demonstrated an individually addressable 21-qubit register of highly coherent and independent qubits with coherence times of about 40~s using nuclear spin qubits in optical tweezers, opening the gate to intermediate scale quantum memories.

\subsubsection{Quantum repeaters}
\label{ch6:repeaters}
As we have previously discussed, light is the most natural long-distance carrier of quantum states. However, the absorption of light imposes intrinsic physical limits on the distance at which single photons can travel. In long-distance fiber communications, absorption is mainly produced by the fiber, with an attenuation coefficient in the range $\sim0.14-0.4$~dB/km in low loss telecom fibers~\cite{G652,Tamura2018}. Furthermore, even in the short-distance communication range of a datacenter, the rate at which photons are lost is nontrivial: the typical loss per SC connector is $\sim0.25$~dB \cite{Sugita1989}, so the shortest possible connection between two nodes accounts for $\sim0.5$~dB of attenuation, i.e., $\sim11$\% of the photons are lost. Hence, if frequent quantum communication is required for a distributed quantum algorithm, the error probability quickly increases as $e=1-10^{n\cdot {\rm dB}/10}$ after $n$ exchanges, limiting the scalability and reliability of the calculation.

It is important to understand that any improvements in the connector losses and fiber attenuation cannot and will not solve the problem of exponential decay with $n$. Given that standard telecommunications \gls*{edfa} cannot be used to amplify arbitrary quantum states due to the no-cloning theorem, {\it quantum repeaters} are essential to the implementation of entanglement distribution and teleportation which enable deterministic transmission of quantum states and remote quantum operations between nodes~\cite{Yoshihisa1994, Takeoka2014}. An early solution to the problem of implementing a quantum repeater was proposed by Briegel et al.\ \cite{Briegel1998}, which consisted of first entangling noisy and imperfect qubits to then create a high-fidelity entangled pair through entanglement distillation. Recent proposals have extended the idea of entanglement distillation to qudits (i.e., $d$-state systems)~\cite{Miguel-Ramiro2018}, multiple simultaneously entangled degrees of freedom (hyperentanglement)~\cite{Hu2021, Du2023}, and logical qubits~\cite{Zhou2016, Luo2022}. Van Leent et al.\ \cite{vanLeent2022} have demonstrated single-atom entanglement over a 33~km telecom fiber using quantum repeaters, proving that long distance entanglement is already a technical possibility. Recent work has also shown that Er$^{3+}$ inclusions in calcium tungstate greatly diminish optical spectral diffusion~\cite{Ourari2023}, a requirement to generate indistinguishable single photons needed for optical repeaters, as this ion is well coupled by its telecom band optical transition. 

\begin{figure*}[t]
    \centering
    \newlength{\twosubht}
    \newsavebox{\twosubbox}
    \savebox\twosubbox{
      \resizebox{0.85\textwidth}{!}{
        \includegraphics[height=3cm]{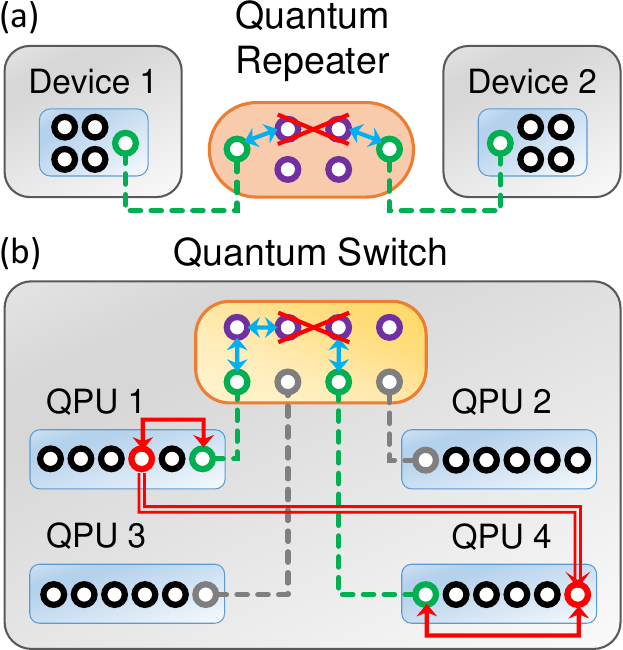}
        \includegraphics[height=3cm]{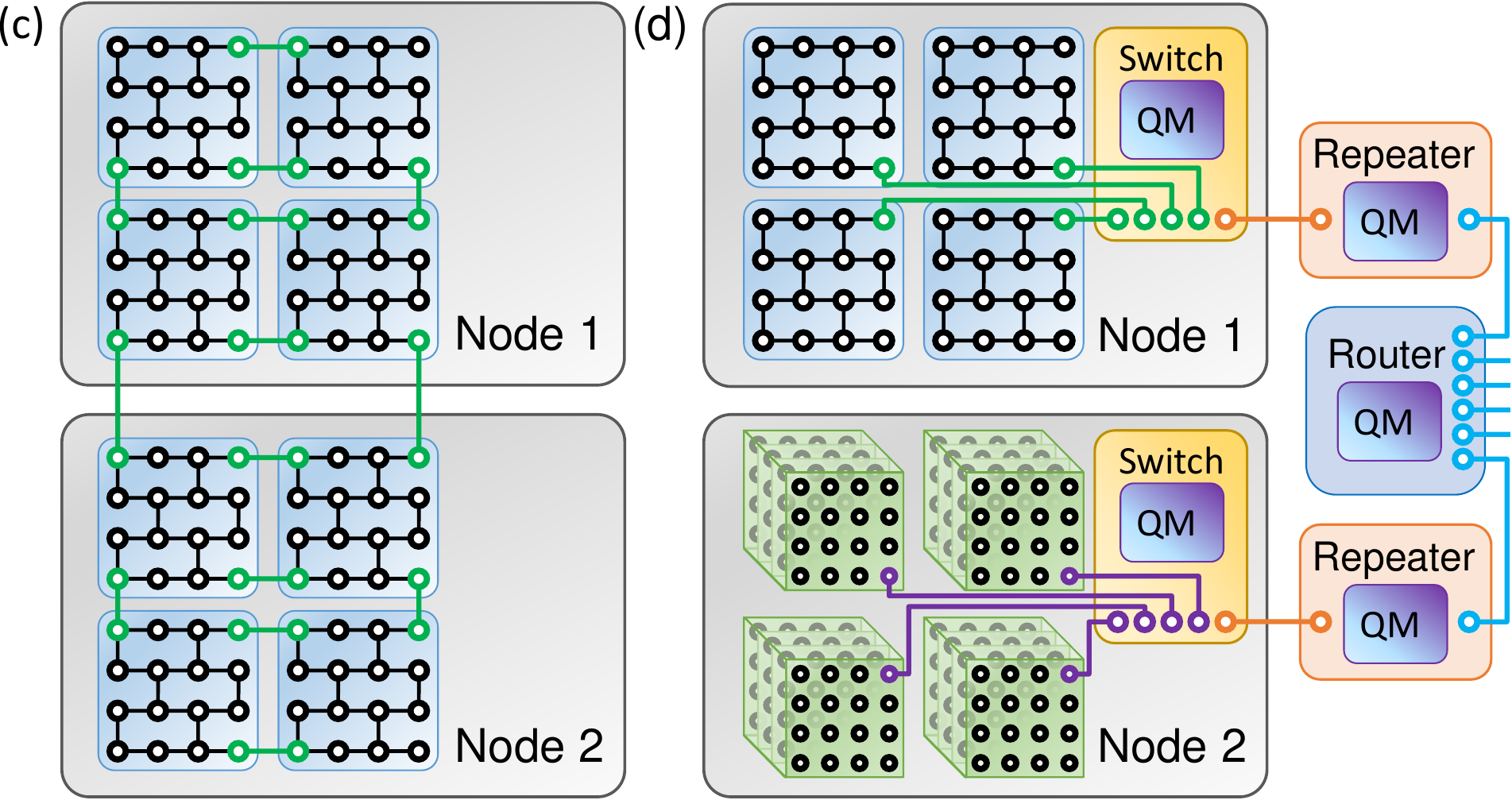}
      }}
    \setlength{\twosubht}{\ht\twosubbox}
    \includegraphics[height=\twosubht]{Images_chapter7_network-a.pdf}
    \includegraphics[height=\twosubht]{Images_chapter7_new-network-b.pdf}
    \caption{Quantum networking devices and interconnects for distributed quantum computing (see main text for details).}
    \label{fig:network-topology}
\end{figure*}

Fig.~\ref{fig:network-topology} (a) shows a schematic representation of a quantum repeater connecting two arbitrary quantum devices. In this figure, qubits are represented as circles, and links are shown as lines, with different colors hinting at the different technologies (e.g., phononic, photonic, or electronic) or energy ranges (e.g., microwave, infrared) used to interconnect the quantum devices, coupled with adequate transducers. Distilled qubits can then be stored in a registry through swap operations (shown as blue arrows) to produce entanglement between the two end devices by performing a \gls*{bsm} between the registry qubits (shown as crossed qubits in red), freeing up the registry qubits and effectively entangling the communication qubits of both devices (shown as green circles).

\subsubsection{Entanglement routers and switches}
\label{ch6:routers-switches}
As previously explained, the execution of general quantum algorithms in multiple qubit-limited \glspl*{qpu} requires entanglement to be generated on demand between pairs of arbitrary qubits~\cite{Laurat2018}. For this reason, recent research has focused on implementing teleportation protocols between non-neighboring nodes. The simplest way to obtain arbitrary entanglement with interconnected \glspl*{qpu} is pre-establishing shared entanglement, as discussed in Section \ref{ch6:transducers}, in a \textit{one-to-one} fashion between specific communication qubits in different nodes. In these \textit{one-to-one} schemes, not every pair of \glspl*{qpu} ought to be physically connected, reducing the complexity of implementation for small integrated systems. 

However, this apparent simplicity suffers from a high scalability burden, leading to significant qubit swap and distillation overhead in complex, strongly entangled algorithms~\cite{Cirac1999}. Even though compilation optimizations can reduce the number of swap operations, more general and modular quantum networks will require {\it entanglement routers} and {\it switches} that will tackle the problem of distributing entanglement between arbitrary qubits, analogous to their classical counterparts~\cite{Zhu2023,Cacciapuoti2023,Zeng2023}. 

For quick reference, classical routers are capable of finding optimal routes in a complex network and understand the \gls*{ip}, while switches only recognize which physical addresses are routed through their connections to redirect traffic. The current absence of a quantum \gls*{ip} standard makes the distinction of the quantum counterparts difficult, so authors have been using these terms interchangeably. Moreover, the quantum hardware required is essentially the same and any differences would arise from the higher-level classical network management. Following this description, any two \glspl*{qpu} in the network can be connected through either one or multiple switches and/or routers in a \gls*{qlan}, or through an efficient routing path that connects multiple routers (which may require repeaters to maintain entanglement) and lead to a \gls*{qwan}~\cite{Renger2023,Gyongyosi22}. The interconnection of quantum networks could eventually lead to a worldwide Quantum Internet. However, this escapes the scope of this review~\cite{Gyongyosi22,Pant2019}. 

Entanglement switches and routers can then be thought of as single-purpose \glspl*{qpu}: their sole objective is establishing entanglement among compute nodes through entanglement swapping, for which implement all the quantum technology required, such as quantum registries, entanglement sources and means to perform \gls*{bsm}, as well as all the hardware required for networking logic and classical communications~\cite{Pant2019}. Moreover, these devices may also be built on different quantum platforms than the proper \glspl*{qpu}, e.g., not requiring the implementation of a complete set of quantum gates but only those required for the swapping protocol and instead requiring registries of qubits with very high fidelity and coherence times longer than the entanglement distillation protocol, or access to quantum memories that fulfill these two requirements. Some proposals suggest networks based on single atoms trapped and coupled to optical resonators as memory qubits, which have long coherence times and good photon coupling (see~\cite{Reiserer2015} and references therein). An example schematic of a quantum switch is shown in Figure \ref{fig:network-topology}(b), where, similarly to quantum repeaters, the distilled qubits are stored in a registry, which can then be used to perform a \gls*{bsm} to entangle any two of the connected devices (shown as \glspl*{qpu} on the drawing) on demand. When entanglement has been distributed, the teleportation protocol can take place (shown as red arrows in (b)). Figs.~\ref{fig:network-topology}(c) and \ref{fig:network-topology}(d) show two examples of \gls*{qlan} architectures, following \textit{one-to-one} and modular topologies respectively. A \textit{one-to-one} topology may be sufficient for smaller systems. However, a more traditional network structure becomes necessary as connectivity grows to tens or hundreds of \glspl*{qpu} in multiple nodes. As each device of the quantum network may have different desirable features and transducers, the hierarchy of a modular network improves scalability and interoperability and unlocks additional performance by offloading overhead to single-purpose entanglement distribution hardware.

%% file: Section3.tex
\section{Networks for distributed quantum computing}
\label{sec:networking}

The previous section describes the various quantum technologies for the implementation of \gls*{dqc}, such as telegate and teledata. However, the implementation of any of these mechanisms will require the physical connection between \glspl*{qpu} to use basic network architectures, such as \textit{point-to-point} or bus, or more complex ones, such as \glspl*{qlan} or \glspl*{qwan}, and the establishment and distribution of entanglement among the \glspl*{qpu}. Nevertheless, classical network architectures and protocols cannot be directly extrapolated to quantum networks for entanglement distribution due to their particularities compared to the transmission of classical bits, such as:
\begin{itemize}
    \item The duration of entanglement mechanisms, and the lifetime of the qubits and the storage time of the qubits in memory due to decoherence.
    \item The probabilistic nature of the different mechanisms, such as the generation of entangled pairs and entanglement swapping.
    \item The need for mechanisms to improve fidelity, such as distillation, both in each independent link and in paths between nodes made up of multiple links.
    \item The possibility of joining entanglement links not only through sequential operations but also through operations carried out in parallel on the various links. The sequential operation is the most similar to the mode of operation of classical networks, in which a data packet goes from source to destination progressively hop by hop.
    \item The different entangled resources --~bipartite, multipartite by means of \gls*{ghz}, W, cluster states and so on.
    \item The need for both quantum and classical channels to achieve the desired functionality. 
    \item The possible use of quantum networks not only for the transmission of quantum information but also for the distribution of entanglement between distant points, which can be used as a resource by itself.
\end{itemize}

\Glspl*{qn} allow the creation and distribution of entanglement between two or more qubits that may be very close to each other or separated by long distances, depending on whether communication takes place between \glspl*{qpu} located on the same node or at geographically distant points. The entanglement resources provided by these \glspl*{qn} will be used both in \gls*{dqc} and other applications of quantum technologies such as sensing, encryption, etc. Li et al.~\cite{net-Li-QI} defined \glspl*{eaqn} as ``network infrastructures formed by interconnecting numerous quantum nodes, which can realize quantum information transmission between arbitrary quantum nodes under the government of network designs and the fundamental laws of quantum mechanics'' \cite{net-ieee-report}. \gls*{dqc} benefits from \glspl*{eaqn} as mechanisms to connect \glspl*{qpu} that are otherwise isolated at the quantum level.

Various works have advanced the research in quantum networks for entanglement distribution, proposing architectures, protocols, and protocol stacks for their implementation in both local and wide area networks. Although a very relevant part of the scientific literature is oriented towards communication systems for the Quantum Internet, they have a common part about entanglement distribution that is relevant to \gls*{dqc}. Particularly, they are suitable for explaining how to connect \glspl*{qpu} in short-distance \glspl*{qlan} or, in other words, how to establish a multi-\gls*{qpu} interconnection among nodes of a datacenter.

A few proposals for quantum network architectures, protocols, and protocol stacks have been summarized and compared in several works \cite{net-ieee-report,net-Qinet-report}. Below are some examples of network proposals, for creating bipartite and multipartite entanglement distribution networks.

\begin{itemize}
    \item Van Meter et al.~\cite{net-vanMeter-Design-QR-nets,net-vanMeter-QI-Arch} propose a \gls*{qrna} describing five layers of network communications that tackle entanglement distribution end to end. Their approach is different from classic networks, as they propose a recursive layer architecture in which swapping and purification functions are repeated to build \textit{end-to-end} entanglement paths from a sequence of links, being entanglement performed at link level. The bottom layers of the protocol architecture are \textit{Physical} and \textit{Link} layers, and they allow the establishment of entanglement at link level (point-to-point). On top of those layers, the \textit{Remote State Composition} and \textit{Error Management layers} are recursive and are continuously repeated performing swapping and purification from entangled links until the system is able to build an end-to-end entangled path. 
    \item Li et al.~\cite{net-Li-QI} and Whener et al.~\cite{net-SWehner.2019} both propose a protocol architecture for quantum networks based on bipartite entanglement where the mission of physical and link layers is the establishment of reliable entanglement, the network layer's goal is the establishment of long distance entanglement, and the transport layer copes with the qubits reliable/deterministic qubits transmission. 
    \item Dür et al.~\cite{net-Pirker-2019} instead propose an architecture and network stack for quantum networks based on multipartite entanglement (\gls*{ghz} graph states) allowing the generation of graph states of any type among clients. This architecture is composed of four layers: physical, connectivity, link, and network. The main difference to the traditional OSI layer architecture relies on the introduction of the connectivity layer, which is responsible for allowing \textit{point-to-point} or \textit{point-to-multipoint} connectivity, as well as error correction and establishment of long-distance links. The link layer allows the creation of graph states in the network that clients will subsequently use for the creation of end-to-end graph states. 
    
\end{itemize}

The study \cite{net-ieee-report} summarizes several examples of network protocol stack proposals for the case of quantum networks and the comparison with the classical protocol stack, based on the OSI or TCP/IP models. 
Also noteworthy is the publication of the \gls*{irtf} \cite{net-rfc9340} describing the Architectural Principles for a Quantum Internet that gathers a relevant part of the information mentioned above. 

Another important factor in the design of quantum networks commented in the mentioned works is the resource reservation strategy. One aspect concerns the entanglement resource reservation, analog to the classic \textit{connection-oriented} or \textit{connection-less} strategies in classical networks. In the first case, a path is obtained between sender and receiver --~in the case of point-to-point entanglement~-- and the necessary resour\-ces for entanglement are reserved in all links of the path between them. In the second case, entanglement links are created in the various links of the path, and any client can use these resources without resource reservation. Another aspect is related to the memory distribution inside the devices. The resource reservation strategy impacts the network architecture and protocols design.

One final comment is that there are still few proposals about the architecture design,  the technology implementation, the services offered, as well as the mechanisms for error correction with the aim of fault tolerant network devices. There are diverse approaches that take into account all optical to hybrid (quantum dots together with optical for instance), DV vs CV, bipartite or multipartite entanglement resources, etc ~\cite{net-vanMeter-Design-QR-nets,net-Pirker-2019,net-GKP-all-optical}.

\subsection{Classical communications}

\begin{figure}
    \centering
    \includegraphics[width=0.8\linewidth]{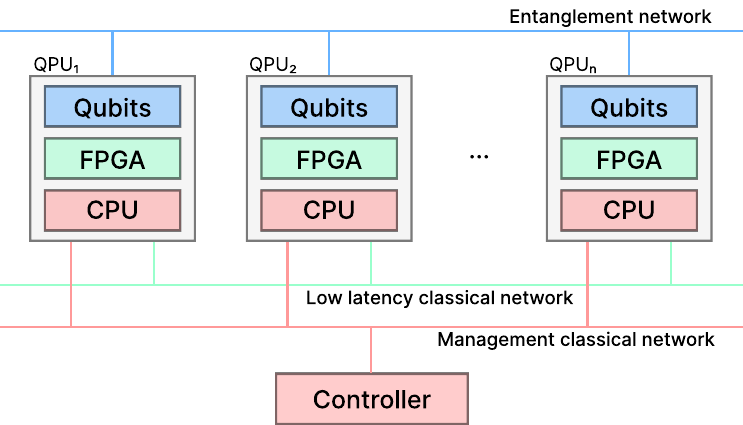}
    \caption{Distributed Quantum Computer Architecture~\cite{net-DiAdamo2021}.}
    \label{fig:net-dqc-arch}
\end{figure}

Classical communications are usually cited but not deeply analyzed in the reviewed literature. To this purpose, a \gls*{dqc} architecture that includes a description of the quantum and classical communications required among elements was proposed by DiAdamo et al.~\cite{net-DiAdamo2021}. Fig.~\ref{fig:net-dqc-arch} depicts the proposed architecture, specifically for short distance connection among \glspl*{qpu}. In this figure, each QPU is defined as a three-layer structure comprising the qubit layer, an FPGA layer for qubit control and measurements, and a CPU layer that is in charge of instructing the FPGA and that includes the interfaces with the \textit{Management classical network}. This management classical network connects the \glspl*{qpu} to the centralized \textit{Controller}. Moreover, the system requires that all nodes are timely synchronized and respond to events on assigned time slots, allowing the scheduling of the execution of each layer of the circuit. This network could be implemented with standard LAN technologies using TCP/IP and/or using an industrial master/slave messaging protocol like Modbus \cite{net-Modbus}. The clock synchronization is implemented among the nodes by means of technologies like \textit{White Rabbit} \cite{net-WhiteRabbit} that could be integrated into the central controller. Finally, a direct \textit{entanglement network} and \textit{low latency classical network} among QPUs for the execution of non-local gates is also suggested. The classical communication is direct between FPGAs not traversing the CPUs of the QPUs. For this classical communication, the authors propose the use of 10~Gbps ethernet LAN technology or industrial protocols for secure, reliable low latency communications, i.e., \textit{Mirrored Bits} \cite{net-MirroredBits}.
 
%% file: Section4.tex
\section{Development layer}
\label{compiler:sec:distributed-quantum-compilers}
\begin{figure}[t]
    \centering
    \includegraphics[width=0.6\linewidth]{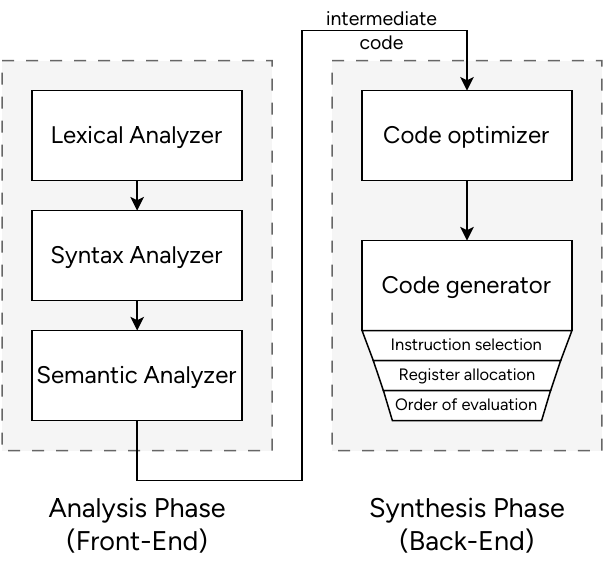}
    \caption{Sequential phases of classic compiler process: analysis and synthesis stages.}
    \label{compiler:fig:compiler-scheme}
\end{figure}

In the realm of classical computing, compilation serves two primary purposes: translating complex programming constructs into machine-specific executable instructions and optimizing machine resources to produce efficient code. Typically, this process follows a common scheme, as illustrated in Fig. \ref{compiler:fig:compiler-scheme}, which consists of two main phases: analysis and synthesis. The analysis phase is responsible for conducting the code's lexical, syntactic, and semantic analysis to ensure correctness. Once validated, the code is translated into an \gls*{ir}, which simplifies the implementation of optimizations in the synthesis phase.

Regarding quantum compilation, the scheme followed is usually the same as in the classical world. This is mostly because quantum compilation turns out to be a fully classical task, leaving the quantum workload just for the execution part. This leads to the situation where many quantum development software tools are actually built on top of classical languages, allowing the analysis phase to be integrated into an existing implementation.

Adding distribution to this task does not alter the compilation scheme; it remains largely the same with some additional steps and restrictions. To fully picture the differences and intricacies of compiling a distributed program, this section will be divided into two parts: Sec.\ \ref{compiler:subsec:types-of-distribution} will elucidate the various methods by which a quantum process --~usually referred to as a quantum circuit~-- can be distributed, while Sec.\ \ref{compiler:subsec:compilation} will delve into how the compilation process is executed considering the distributed nature of the task.

\subsection{Types of distribution}
\label{compiler:subsec:types-of-distribution}
Distributed computing makes it possible to organize the computation of a problem in different \acp*{pu}, which are connected through an interconnection network. The advantages of this model are evident: reducing the execution time by leveraging multiple \glspl*{pu} computing in parallel or, for large problems that do not fit within a single node, partitioning them to enable their solution. The time reduction comes with its own set of disadvantages, notably the increased difficulty in adapting algorithms and codes to a distributed approach. This is due to the significant overhead caused by communications and synchronizations, which must be carefully considered and managed~\cite{Amza1996}.

Therefore, the complexity of developing a code increases when it is distributed. This complexity especially impacts the compiler design. In the analysis phase, new communication directives need to be developed, while in the synthesis phase, various network architectures must be considered to optimize data transmission and reception~\cite{Van2017}. 

\begin{figure}[t]
    \centering
    \includegraphics[width=0.8\linewidth]{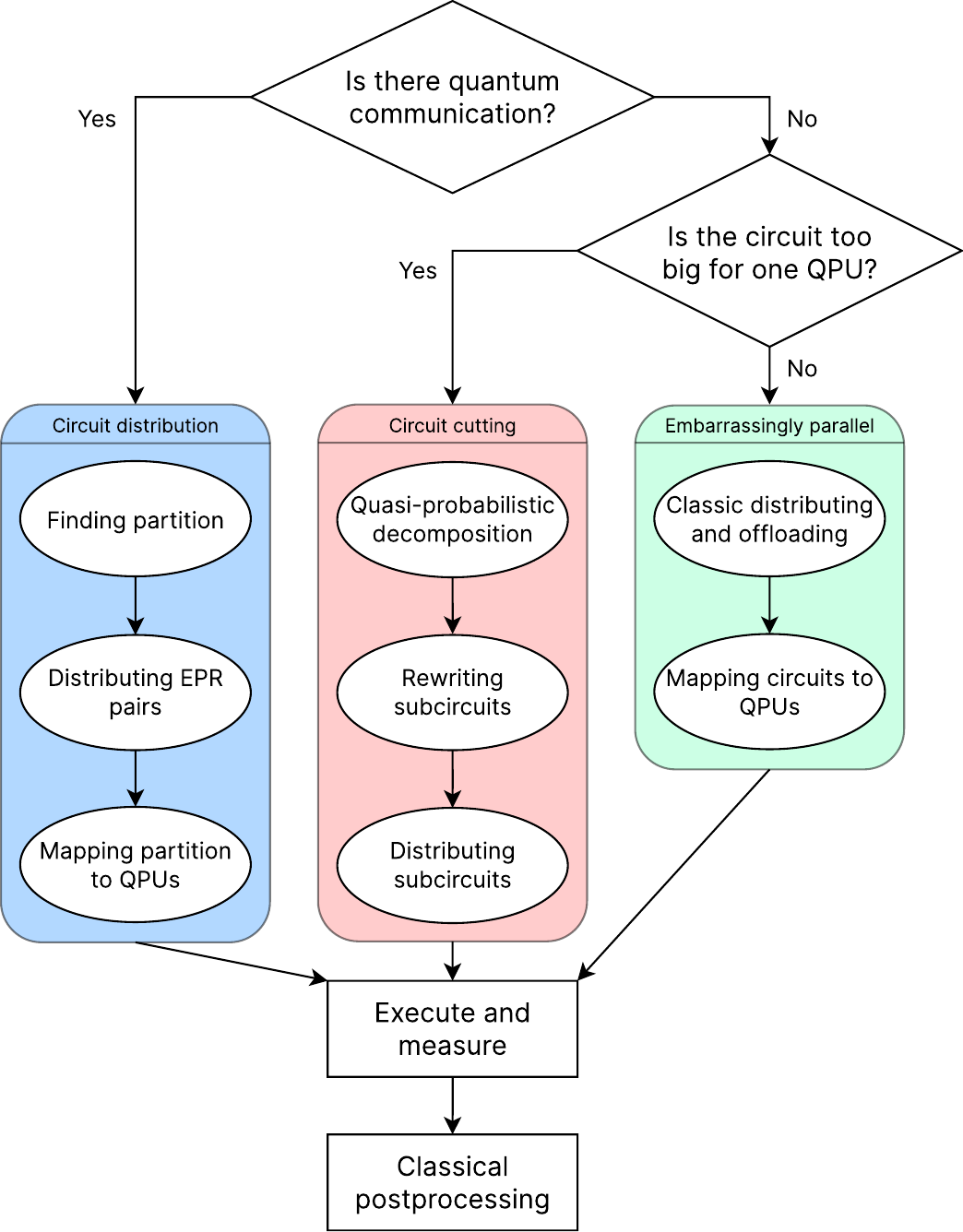}
    \caption{Types of quantum distribution and their stages simplified.}
    \label{compiler:fig:distribution-types}
\end{figure}

Certainly, the network's communication mechanisms and the resources required by the quantum task dictate the applicable distribution model, as depicted in Fig. \ref{compiler:fig:distribution-types}. Three distinct categories of quantum distribution emerge: \textit{circuit distribution}, \textit{circuit cutting}, and \textit{embarrassingly parallel}. It is clear, looking at Fig.~\ref{compiler:fig:distribution-types}, that all categories converge in executing, measuring, and post-processing information. Now, we will elucidate the stages where each distribution type diverges.

First of all, \textit{circuit distribution} is associated with the existence of a quantum communication network --~assuming the existence of a classical network as shown in Fig.\ \ref{fig:net-dqc-arch}. This capability permits the execution of a single circuit that demands more qubits than available in a single \gls*{qpu}. In this case, the steps involved are:

\begin{enumerate}
    \item Finding the partition. This stage is responsible for defining how the quantum circuit is going to be distributed among the \glspl*{qpu}. Nevertheless, determining the partition of a quantum circuit is a non-trivial task, as finding an optimal or near-optimal solution is complex. While some software tools exist to perform this task, it remains challenging.
    \item Distributing \gls*{epr} pairs. To enable circuit distribution, quantum communication resources need to be established, which involves generating entanglement between pairs of arbitrary qubits. This process is directly linked to the generation of entanglement on demand between interconnected quantum processing units (\glspl*{qpu}) discussed in section \ref{ch6:routers-switches}. 
    \item Mapping partition to \glspl*{qpu}. Once the circuit is partitioned and the quantum communication resources are available, the circuit is mapped to the physical structure. This involves a local mapping in each of the \glspl*{qpu} along with the establishment of the quantum communication operations necessary, as explained in section \ref{ch6:quantum-teleportation}.
\end{enumerate}

Alternatively, if quantum communication is not available and the circuit is too large to fit in a single \gls*{qpu}, \textit{circuit cutting} may be employed. Similar to circuit distribution, we assume that classical communication is always available to allow nodes to share their results during the post-processing stage. Now, the steps to perform circuit cutting are as follows: 

\begin{enumerate}
    \item Finding the partition. This is an analogous stage to the circuit partitioning in circuit distribution. A partition that minimizes the number of \gls*{epr} pairs will also minimize the classical cost incurred in circuit cutting. The extra classical cost of circuit cutting becomes exponential with the number of \gls*{epr} pairs that would be needed in the fully distributed protocol.
    \item \Gls*{qpd}. Since quantum communication resources are not available, it has to be simulated classically. The circuit is divided into subcircuits to be executed independently on each available \glspl*{qpu}. Each of these subcircuits has an associated weight in the \Gls*{qpd} given by an appropriate decomposition of the original circuit in the partitions, and the final outcome of the computation is recovered as the weighted combination of the outcomes of the subcircuits. Crucially, these weights can be either positive or negative, hence the quasi-probability, and the number of subcircuits grows exponentially with the amount of quantum communication to be simulated. 
    \item Distributing the subcircuits. Finally, each subcircuit is scheduled for execution on a specific \gls*{qpu}, and a local mapping is performed before execution takes place.
\end{enumerate}

Finally, if no quantum communication is available and the circuit fits in one \gls*{qpu}, then the technique to apply might be \textit{embarrassing parallelism}. The steps required in this case are:

\begin{enumerate}
    \item Classic distributing and offloading. Classic distribution means that each \gls*{qpu} is scheduled to execute a determined part of the quantum task, distributing in that way the workload. On the other hand, classic offloading refers to the execution of a classic program with some quantum tasks that are \textit{offloaded} to a corresponding \gls*{qpu}.
    \item Mapping the circuits to \glspl*{qpu}. As before, once the classic distributing or offloading is performed, circuits are mapped to the corresponding \gls*{qpu}.
\end{enumerate}

It is important to remark that these distribution types are not mutually exclusive, but quantum compilers typically select one option. The closest work to combine several distribution types is the one by Tomesh et al.\ \cite{Tomesh2021DivideComputation}. They introduced the \gls*{qdca}, a hybrid variational approach aimed at mapping large combinatorial optimization problems onto distributed quantum architectures. This was accomplished by leveraging graph partition and circuit-cutting techniques in combination. We will delve more into it in section \ref{compiler:subsubsec:compilers}.

Now, each of the groups that have just been outlined will be dissected to fully understand how the quantum distribution works in each case. First, we will look at circuit distribution --~the most common type of distribution~-- in section~\ref{compiler:subsubsec:circuit-distribution}. Techniques for circuit cutting are analyzed in section~\ref{compiler:subsubsec:circuit-cutting}. Finally, in section~\ref{compiler:subsubsec:embarrasingly-parallel}, solutions for embarrassingly parallel problems are presented.

\subsubsection{Circuit distribution}
\label{compiler:subsubsec:circuit-distribution}

Circuit distribution, as has been presented, involves three main phases: first, finding an optimal or near-optimal partition; second, distributing the partition among the available \glspl*{qpu}, and third, mapping this partition to each \gls*{qpu}. However, partitioning the circuit presents the most significant challenge and will be the primary focus of our efforts in this section. The other aspects are common to all the distribution types and will be further explained in the compilation section \ref{compiler:subsec:compilation}.

First, for partitioning, the quantum circuit is mapped onto a graph that shows interconnections between elements. Thus, quantum circuit partitioning turns into a graph partitioning problem: given an undirected graph $G=(V, E)$ with a vertex set $V$ and an edge set $E$, the aim is to partition $V$ into two or more subsets regarding a cost function, like the number of edge cuts generated by the partition.  

\begin{figure}[t]
    \centering
    \includegraphics[width=0.5\linewidth]{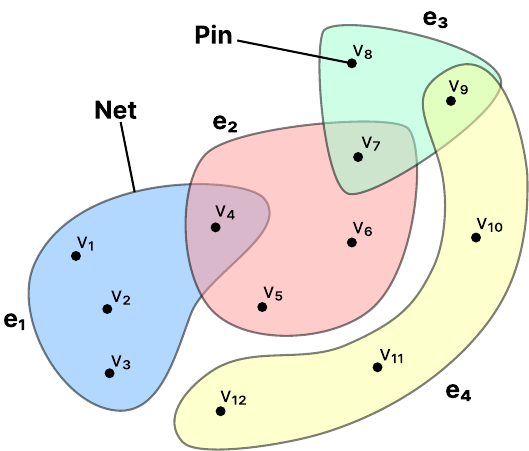}
    \caption{Example of a hypergraph with twelve pins ${v_i}$ and four nets ${e_j}$. Net $e_1$ has a size of four as it ensembles four pins, and pin $v_4$ has a degree of 2 as it belongs to two nets.}
    \label{fig:hypergraph}
\end{figure}

Graphs assume that the interaction between vertices is by pairs. However, even the most trivial phenomenon implies more vertices interacting concurrently. It is necessary to broaden the graph concept to gather these multilateral connections. The so-called \textit{hypergraphs} \cite{berge1984hypergraphs} generalize the graphs to more complex situations. In short, while a graph can establish connections by pairs, a hypergraph is an object that connects more than two vertices or pins through elements called hyperedges or nets, as shown in Fig.~\ref{fig:hypergraph}. Thus, a hypergraph $H=(V, E)$ is an ensemble of pins $V$ and nets $E$ among those pins, and a net $e\in E$ is a subset of more than two pins.

\begin{table}[t]
\centering
\footnotesize
\begin{tabular}{|l|l|}
\hline \hline
\textbf{Hypergraph partitioning} & \textbf{Circuit distribution} \\ \hline \hline
Vertices                & Wires (qubits)       \\ \hline
Hyperedges              & Groups of CZs        \\ \hline
Partition               & Distribution         \\ \hline
Blocks                  & \glspl*{qpu}         \\ \hline
\end{tabular}
\caption{Translation of hypergraph partitioning to circuit distribution extracted from the original paper \cite{Andres-Martinez2019AutomatedPartitioning}.}
\label{table:Hypergraph}
\end{table}

Hence, hypergraph partitioning generalizes graph partitioning. More precisely, a $k$-way hypergraph partitioning groups the pins of a hypergraph into $k$ blocks minimizing an objective function so that few nets connect pins from different blocks. The exchangeable objective functions are the cut-net and the connectivity metrics. The cut-net metric generates independent blocks of vertex sets by minimizing the nets belonging to several blocks, whereas the connectivity metric weights each net $e$ with a factor $\lambda_e-1$ to diminish the $\lambda_e$ blocks connected by a net. The cut-net objective function sums over the nets among blocks and the connectivity metric over the $\lambda_e$ blocks connected by a net. Nevertheless, both are analog to the edge-cut problem in graph partitioning.

Underneath the goal of minimizing the cut-net and connectivity metrics lies an important consideration: while a valid partition may suffice for \gls*{dqc}, it may not necessarily be an optimal partition. For instance, in the circuits responsible for teledata and telegate operations --~as illustrated in Fig. \ref{fig:circuits_teledata_telegate}~--, these operations add up to four layers of depth to the circuit to enable operations among qubits in different \glspl*{qpu}. Consequently, this introduces latency to the quantum circuit, especially considering the additional synchronization required for intermediate measurements contained in both protocols between both \glspl*{qpu}. This latency represents a significant bottleneck in circuit distribution. Therefore, all circuit partitioning methods aim to minimize the utilization of teledata or telegate protocols. This aspect will be crucial in the circuit distribution techniques discussed in this section and beyond.

In order to realize this partitioning, some classical algorithms are usually employed as a third-party algorithm. Two of the most common are \gls*{kahypar}~\cite{akhremtsev2017engineering, schlag2023high} and \gls*{kl}~\cite{kernighan1970efficient}. \gls*{kahypar} is a multilevel hypergraph partitioning framework that enhances net cut and connectivity metrics. \gls*{kahypar} utilizes coarsening and port\-fo\-lio-based initial partitioning. First, \gls*{kahypar} applies coarsening for grouping the pins into nets, reducing the number of pins. Second, when the number of nets is small enough, \gls*{kahypar} employs portfolio-based initial partitioning that compares results from several optimizers and selects the best, enhancing the partitioning power. Finally, an uncoarsening process returns the partition of the original hypergraph. The running time is linear $\mathbb{O}(n)$ in the number $n$ of gates. Similar to the uncoarsening step of \gls*{kahypar}, the \gls*{kl} algorithm is a heuristic algorithm for graph partitioning to divide the graph vertices into two subsets to reduce the edges across the subsets. Of course, these are not the only algorithms or models. In another approach, proposed by Clark et al.\ \cite{clark2023TDAG}, a different model than hypergraph is employed. They introduce the \gls*{tdag} partitioning for quantum circuits, a novel method that views circuits as a series of binary trees and selects the tree containing the most gates for partitioning.

Two of the first approaches aiming to reduce communication between partitions are the work of Zomorodi et al.\ \cite{Zomorodi-Moghadam2018OptimizingCircuits} and of Martínez and Heunen \cite{Andres-Martinez2019AutomatedPartitioning}. The former is a special case where only two \glspl*{qpu} is considered. They use the \gls*{kl} algorithm as used in the VLSI design algorithms to minimize communication between the two partitions. After that, they apply a custom algorithm which aims to reduce the number of teleportations applied. The latter, by Martinez and Heunen, is one of the most significant contributions in the field, serving as a foundational reference in many of the articles discussed here. Their method involves two key phases: a pre-processing phase, which groups equivalent gates, and a second phase, where hypergraph partitioning is performed using \gls*{kahypar}. They evaluated their algorithm using five quantum algorithms known for their quantum speedup, such as \gls*{qft}.

The Zomorodi et al.\ work was later improved by Houshmand et al.\ \cite{Houshmand2020AnComputation} by exchanging the algorithm responsible for reducing teleportations --~which had exponential cost~-- for a genetic one. They achieved similar results with a significant decrease in execution time. However, they criticized the work of Martinez and Heunen for not considering optimizations such as moving gates back and forth to bring them closer together, as proposed by Zomorodi et al. in their work. Additionally, Martinez and Heunen did not explore the entire search space of different partitioning options for executing global gates, which limited their ability to produce optimal solutions. But these two approaches \cite{Zomorodi-Moghadam2018OptimizingCircuits, Houshmand2020AnComputation} only consider a two \gls*{qpu} scheme, reason why Daei et al.\ \cite{daei2020optimized} enhanced it by effectively mapping a quantum circuit into an appropriate number of distributed components. Moreover, Nikahd et al.\ \cite{Nikahd2021AutomatedCircuits} also took a step further categorizing the binary gates into distinct ``levels'', followed by determining the optimal partitioning of qubits for each level through the solution of an integer linear program.  

The work by Martinez and Heunen \cite{Andres-Martinez2019AutomatedPartitioning}, on the other hand, was extended with an entanglement-ef\-fi\-cient protocol \cite{wu2023entanglement} derived from \cite{eisert2000optimal} and with, among other things, a hypergraph approach to arbitrary network topologies \cite{andres2023distributing}. In the first case, authors pack multiple non-local controlled unitary gates locally with one maximally entangled pair through a distributing and embedding pipeline. In the second, the authors also search for efficient entanglement within the network by reusing already available connections. In fact, this work led to many different articles employing hypergraph partitioning with \gls*{kahypar}, as shown in this section. 

Another work that employed \gls*{kahypar} was developed by Sundaram et al.\ \cite{sundaram2021} which presents a two-step heuristic for the distribution of quantum circuits: dividing the given circuit’s qubits among the computers in the network --~where the \gls*{kahypar} algorithm is employed~-- and scheduling communication operations, called migrations --~equivalent to cat-entanglement operations \cite{Yimsiriwattana2004Generalized}. They present a polynomial-time solution for the second step in a special setting and a $\mathbb{O}(\log n)$-approximate solution in the general setting. The same authors improved the work by amplifying the available remote protocol employed \cite{Sundaram2022DistributionNetworks}.  While Daei et al.\ \cite{daei2020optimized} use teledata as the only means of communication between \glspl*{qpu} and, on the contrary, Martinez and Heunen \cite{Andres-Martinez2019AutomatedPartitioning} and Sundaram et al.\ \cite{sundaram2021} use telegate, this work employs both. For the latter, i.e., the telegate protocol, they used a method similar to the improved work with a two-step heuristic. Notwithstanding, they used a Tabu-search-based heuristic to partition the given circuit’s qubits among \glspl*{qpu}, considering the network's heterogeneity and the storage limits. And for the general \gls*{dqc} problem they employed two heuristics: \textit{Sequence}, a greedy approach, and \textit{Split}, similar to the previous one, but with an iterative approach. Both employ the telegate solution as a subroutine. Even more, Sundaram et al.\ took a step further in a recent work \cite{Sundaram2023DistributingTeleportations} by designing two different protocols aiming to reduce the number of teleportations needed to perform the distributed task. The first method, termed Local-Best, tries to minimize the teleportation of qubits by selecting them only when necessary, with the choice of teleportation being influenced by gates in the near future. The algorithm consists of two steps:

\begin{enumerate}
    \item Find an initial assignment of qubits to computers to minimize the number of resulting non-local binary gates.
    \item For each non-local binary gate G, select the teleportations to execute G locally based on the ``near future'' in order to minimize the total number of teleportations.
\end{enumerate}

The second method, named Zero-Stitching, comprises two main steps:

\begin{enumerate}
    \item Identify ``zero-cost'' subcircuits: These are contiguous subcircuits that can be executed without any teleportations.
    \item Divide the given circuit into zero-cost subcircuits and ``stitching'' them together using teleportations.
\end{enumerate}

There were also approaches employing bipartite graphs instead of hypergraphs. Davarzani et al.\ \cite{davarzani2020dynamic} proposed an algorithm for distributing quantum circuits to optimize the number of teleportations between qubits that consisted of two steps: first, the quantum circuit was converted to a bipartite graph (bigraph), and, second, the bigraph was partitioned into $K$ parts employing a dynamic programming approach. Finally, they compared their results with the ones yielded by works previously analyzed \cite{Andres-Martinez2019AutomatedPartitioning, Houshmand2020AnComputation, Zomorodi-Moghadam2018OptimizingCircuits} and they claimed that the experiments gave better or equal results for benchmark circuits.

Besides minimizing the communication between partitions, in \cite{cambiucci2023hypergraphic} adjustable scenarios to the capabilities and constraints of the processing units involved in the distribution are considered. In this work, instead of the \gls*{kl} from the original hypergraphic approach, authors implement a variation of the Fiduccia-Mattheyses algorithm \cite{fiduccia1988linear}, which is a faster approximation algorithm for min-cut partitioning with a computational time that grows linearly with the network size. They use the same circuits as \cite{Andres-Martinez2019AutomatedPartitioning} for benchmarking.

A field-changing approach was the work developed by Baker et al.\ \cite{Baker2020Time-slicedArchitectures}. While still based on graph partitioning, this method seeks to avoid reaching a single static assignment for an entire circuit by employing near-optimal graph partitioning techniques. It leverages the inherent clustering of the \gls*{dqc} paradigm and the statically-known control flow of quantum programs to develop tractable partitioning heuristics. These heuristics map quantum circuits to modular physical machines one time slice at a time. Specifically, optimized mappings are created for each time slice, considering the cost to move data from the previous time slice and utilizing a tunable lookahead scheme to reduce the cost of moving to future time slices. To achieve this, a customized version of the \gls*{oee} algorithm \cite{park1995algorithms} --~considered a natural extension of the \gls*{kl} algorithm~-- referred to as \gls*{roee}, is employed. Because the primary approach to map the circuit to the hardware is \gls*{fgp}, this method is usually referred to as \gls*{fgp}-\gls*{roee}. This method was further analyzed by Ovide et al.\ examining it under another multi-core architecture but maintaining the all-to-all qubit and cores connectivity \cite{ovide2023mapping}. Moreover, a \gls*{hqa} method for partitioning is developed by Escofet et al., which also describes the assignment of qubits to cores between timeslices, and it is compared to the \gls*{fgp}-\gls*{roee} method \cite{escofet2023hungarian}.

A recent approach that has elevated the work of Baker et al.\ is the technique presented by Bandic et al.\, which employs a \gls*{qubo} approach in order to partition the circuit at each time slice \cite{bandic2023mapping}. Their method's primary strengths are rooted in the formulation of the \gls*{qubo} itself. This structure enables the decoupling of the problem definition from the solver as well as surpassing the limitations of look-ahead approaches utilized in the Baker et al.\ solution. It is worth noticing that, in this approach, two different multi-core architecture layouts composed of 10 cores with a capacity of 10 qubits each were tested, in contrast with the non-realistic all-to-all connectivity assumed by the previous approaches.

Last but not least, one of the most novel algorithms is a circuit partitioning method that employs \gls*{drl} \cite{pastor2024circuit}. Once again, the \gls*{fgp}-\gls*{roee} is employed as a baseline to compare the results and as an inspiration due to its time-sliced graph partitioning. This work has considered three approaches: \gls*{ppo}, Soft Mask, and Hard Mask. The first one, the \gls*{ppo}, is a widely used algorithm within the \gls*{drl} scheme, while the remaining two, Soft and Hard Mask, are a variant of the former \gls*{ppo} algorithm that introduces a masking mechanism. The Soft Mask approach adds a simple mask, which disables useless operations --~such as swapping identical qubits, swapping two qubits situated on the same machine, or advancing to the subsequent time slice without establishing a valid assignment for the current one~-- whereas Hard Mask implements a \textit{direct-swap} heuristic in top of the Soft Mask which solely evaluates the relocation of misplaced qubits to the respective core they need to interact with.

Now that we have explored the state-of-the-art in the circuit partitioning problem, we can understand why it poses such a significant challenge. Finding the optimal partition directly impacts performance and is, therefore, a critical aspect in the later stages of compilation, where the boundaries between software and hardware become narrow. Specifically, this problem is closely related to the qubit mapping and circuit optimization stages of the distributed quantum compiler, which are defined and explained in section \ref{compiler:subsubsec:synthesis-phase} as part of the synthesis phase. We will delve deeper into this link in that section, but it is essential to establish the correlation between performance and the chosen quantum distribution method early on.

\subsubsection{Circuit cutting}
\label{compiler:subsubsec:circuit-cutting}

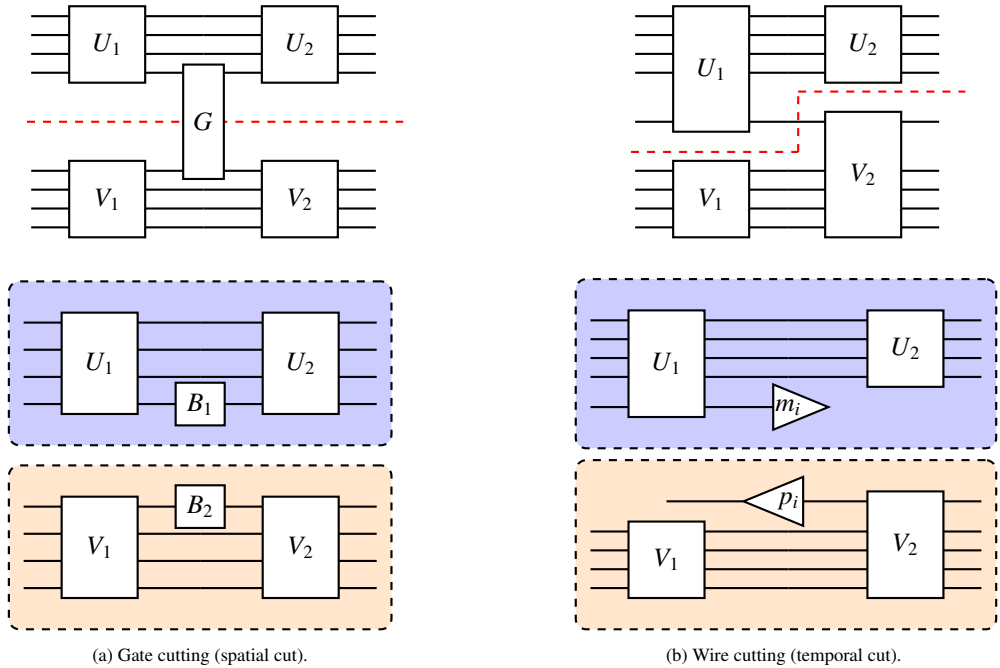
\begin{figure*}[tb]
    \centering
    \begin{subfigure}[b]{0.3\linewidth}
        \centering
        \begin{tikzpicture}[overlay]
            \draw[red, thick, dashed] (0.15,0) -- (5.1,0);
        \end{tikzpicture}
        \begin{quantikz}[row sep={0.25cm,between origins}]
            \qw & \gate[4, style={inner ysep=-0.15cm}][1cm]{U_1} & \qw         & \gate[4, style={inner ysep=-0.15cm}][1cm]{U_2} & \qw \\
            \qw &               & \qw         &               & \qw \\
            \qw &               & \qw         &               & \qw \\
            \qw &               & \gate[2, style={inner ysep=-0.15cm}]{G} &               & \qw \\[1.05cm]
            \qw & \gate[4, style={inner ysep=-0.15cm}][1cm]{V_1} &             & \gate[4, style={inner ysep=-0.15cm}][1cm]{V_2} & \qw \\
            \qw &               & \qw         &               & \qw \\
            \qw &               & \qw         &               & \qw \\
            \qw &               & \qw         &               & \qw
        \end{quantikz}
    \end{subfigure}
    %\hfill
    \hspace{2cm}
    \begin{subfigure}[b]{0.3\linewidth}
        \centering
        \begin{tikzpicture}[overlay]
            \draw[red, thick, dashed] (0.15,-0.4) -- (2.35,-0.4);
            \draw[red, thick, dashed] (2.35,-0.4) -- (2.35, 0.4);
            \draw[red, thick, dashed] (2.35, 0.4) -- (4.55, 0.4);
        \end{tikzpicture}
        \begin{quantikz}[row sep={0.25cm,between origins}]
          \qw & \gate[5, style={inner ysep=-0.15cm}][1cm]{U_1} & \qw & \gate[4, style={inner ysep=-0.15cm}][1cm]{U_2} & \qw \\
          \qw &               & \qw &               & \qw \\
          \qw &               & \qw &               & \qw \\
          \qw &               & \qw &               & \qw \\[0.4cm]
          \qw &               & \qw & \gate[5, style={inner ysep=-0.15cm}][1cm]{V_2} & \qw \\[0.4cm]
          \qw & \gate[4, style={inner ysep=-0.15cm}][1cm]{V_1} & \qw &               & \qw \\
          \qw &               & \qw &               & \qw \\
          \qw &               & \qw &               & \qw \\
          \qw &               & \qw &               & \qw 
        \end{quantikz}
    \end{subfigure}

    \begin{subfigure}[b]{0.3\textwidth}
        \centering
        \begin{quantikz}[row sep={0.36cm,between origins}]
            \qw\gategroup[4,steps=5,style={dashed,rounded corners,fill=blue!20, inner xsep=2pt},background]{} & \gate[4, style={inner ysep=-0.15cm}][1cm]{U_1} & \qw         & \gate[4, style={inner ysep=-0.15cm}][1cm]{U_2} & \qw \\
            \qw &               & \qw         &               & \qw \\
            \qw &               & \qw         &               & \qw \\
            \qw &               & \gate[style={inner ysep=0.001cm}]{B_1} &               & \qw \\[1cm]
            \qw\gategroup[4,steps=5,style={dashed,rounded corners,fill=orange!20, inner xsep=2pt},background]{} & \gate[4, style={inner ysep=-0.15cm}][1cm]{V_1} &  \gate[style={inner ysep=0.001cm}]{B_2}           & \gate[4, style={inner ysep=-0.15cm}][1cm]{V_2} & \qw \\
            \qw &               & \qw         &               & \qw \\
            \qw &               & \qw         &               & \qw \\
            \qw &               & \qw         &               & \qw
        \end{quantikz}
        \caption{Gate cutting (spatial cut).}
    \end{subfigure}
    %\hfill
    \hspace{2cm}
    \begin{subfigure}[b]{0.3\textwidth}
        \centering
         \begin{quantikz}[row sep={0.25cm,between origins}]
          \qw\gategroup[5,steps=5,style={dashed,rounded corners,fill=blue!20, inner xsep=2pt},background]{} & \gate[5, style={inner ysep=-0.15cm}][1cm]{U_1} & \qw & \gate[4, style={inner ysep=-0.15cm}][1cm]{U_2} & \qw \\
          \qw &               & \qw &               & \qw \\
          \qw &               & \qw &               & \qw \\
          \qw &               & \qw &               & \qw \\[0.15cm]
          \qw &               & \gateTriangle{m_i} \\[1cm]
          \gategroup[5,steps=5,style={dashed,rounded corners,fill=orange!20, inner xsep=2pt},background]{} & & \gateTriangleInv{\rotatebox[origin=c]{180}{$p_i$}} & \gate[5, style={inner ysep=-0.15cm}][1cm]{V_2} & \qw \\[0.15cm]
          \qw & \gate[4, style={inner ysep=-0.15cm}][1cm]{V_1} & \qw &               & \qw \\
          \qw &               & \qw &               & \qw \\
          \qw &               & \qw &               & \qw \\
          \qw &               & \qw &               & \qw 
        \end{quantikz}
        \caption{Wire cutting (temporal cut).}
    \end{subfigure}
    
    \caption{Two schemes for cutting a quantum circuit: gate-cutting (or spatial cut) ~\cite{mitarai2021constructing} and wire-cutting (or temporal cut)~\cite{peng2020simulating}. Both can be shown to be equivalent~\cite{Brenner2023optimal}.}
    \label{compiler:fig:cutting-schemes}
\end{figure*}

As detailed in section~\ref{sec:networking}, on the road to fully functional \gls*{dqc}, one needs quantum communication in the form of a quantum network between the devices. In the absence of these kinds of networks, there are several alternative techniques to simulate, or at the very least approximate, this entanglement between parties using a classical network. In this context, circuit cutting has been suggested as a solution to partitioning a wide circuit requiring many qubits into smaller parts with no entanglement. These smaller subcircuits can then be executed (emulated classically) either sequentially in a computer with limited qubits (memory) or in parallel using separate devices. The output of the original circuit is then recovered using a combination of the results of the subcircuits, with some cost in accuracy that has to be overcome by increasing the number of circuit executions as compared to the original. This extra cost is often called sampling overhead. There are several different strategies for circuit cutting, such as gate-cutting and wire-cutting (shown in Fig.~\ref{compiler:fig:cutting-schemes}). Still, in all of them, the sampling overhead is known to grow exponentially with the number of cuts.

\paragraph{Quasi-probabilistic decomposition of quantum channels}
Here, the concept of \gls*{qps} of a quantum circuit is introduced, which is the basis of most forms of circuit cutting, and uses the \gls*{qpd} of the \textit{quantum channel} of the circuit. To understand these, it is helpful to work in the density operator formalism, in which a $n$-qubit quantum state $\rho$ is described by a positive Hermitian matrix of size $2^n \times 2^n$ with trace equal to one. The density operator enables the description of general quantum states, including both pure and mixed states. This formalism allows us to take into account the effect of operations such as intermediate measurements, or the effects of noise (decoherence, dephasing, etc.) using the so-called \textit{quantum channels} (also known as \textit{quantum operations})~\cite{NielsenChuang2010}.

Formally, a quantum channel $\mathcal{E}$ corresponds to a trace-pre\-ser\-ving, completely positive linear map between density operators. The evolution of the initial state $\rho_0$ to the final state $\rho$ is then $\rho = \mathcal{E}(\rho_0)$, and the expected value of an observable $O$ would be
\begin{equation}
    \langle O \rangle = \Tr{O \mathcal{E}(\rho_0)}.
\end{equation}
One usual way of representing general quantum channels is through the operator-sum representation (also known as Kraus decomposition). In this representation, we express the action of the quantum operation $\mathcal{E}$ on a state $\rho$ as a sum of $k$ terms
\begin{equation}
    \mathcal{E}(\rho)=\sum_{j=1}^{k}E_j\rho E^{\dagger}_j \; ,
    \label{compiler:eq:operator-sum}
\end{equation}
where $E_i$ are (Kraus) operators on the Hilbert space of $\rho$. 

The key here is that Eq.~\ref{compiler:eq:operator-sum} is not unique, i.e., one has the freedom to choose the operators $E_i$ of the representation and still get the same channel $\mathcal{E}$. In particular, one can choose the operators to be quantum gates that are \textit{local} in separate sets of qubits. Consider the Hilbert space of our $n$-qubit bipartite system $\rho = \rho^{(1)}\otimes\rho^{(2)}$ as $\mathcal{H}=\mathcal{H}^{(1)}\otimes\mathcal{H}^{(2)}$, where $\mathcal{H}^{(1)}$ and $\mathcal{H}^{(2)}$ are the space of the two sets of qubits $\rho^{(1)}$ and $\rho^{(2)}$, with no physical connection between them. Now consider a quantum circuit $C$ consisting of products of arbitrary quantum gates, some of them multi-qubit gates acting on both $\mathcal{H}^{(1)}$ and $\mathcal{H}^{(2)}$ simultaneously. Our hardware may not be able to execute those non-local gates, but one can always find a decomposition such that
\begin{align}
    \mathcal{E}(\rho) &= \sum_i^m q_i \left( V_i^{(1)}\otimes V_i^{(2)}\right)\left( \rho^{(1)}\otimes \rho^{(2)}\right)\left( V_i^{(1)\dagger}\otimes V_i^{(2)\dagger}\right) \nonumber \\
    &= \sum_i^m q_i \left( V_i^{(1)}\rho^{(1)}V_i^{(1)\dagger}\right)\otimes \left( V_i^{(2)}\rho^{(2)}V_i^{(2)\dagger}\right) \nonumber \\
    &= \sum_i^m q_i \, \mathcal{E}_i^{(1)}\left(\rho^{(1)}\right)\otimes \mathcal{E}_i^{(2)}\left(\rho^{(2)}\right) \; ,
\end{align}
with coefficients $q_i\in\mathbb{R}$ with $\sum_{i=1}^m q_i=1$, and $V_i^{(1)}$ and $V_i^{(2)}$ are operations acting locally in $\mathcal{H}^{(1)}$ and $\mathcal{H}^{(2)}$ respectively, that our hardware can physically execute. The choice of $q_i$ and the set of $V_i^{(1)}$ and $V_i^{(2)}$ is not unique, and it is known as a \gls*{qpd} of the quantum channel~\cite{pashayan2015estimating}.

The $q_i$ can be either positive or negative, which is why they are called quasi-probabilities. The larger the number of negative coefficients in the decomposition, the larger the 1-norm $\kappa = \sum_{i=0}^m |q_i|$ of the \gls*{qpd} becomes. Crucially, this $\kappa$ quantity is related to the cost of executing the circuit $C$ that has non-local gates, using only local operations~\cite{mitarai2021overhead,piveteau2022quasiprobability}. Negative probabilities in the simulation of quantum circuits were already known to be related to the ``quantumness'' of quantum circuits, and thus to how expensive it is to classically simulate quantum processes~\cite{pashayan2015estimating,veitch2012negative,bravyi2016trading,temme2017error}. 

In practice, to calculate the expected value of an observable, we sample the outcome of the circuit measured in the appropriate basis for some number of shots $N_s$. We want $N_s$ to be large enough so as to have some desired degree of accuracy $\epsilon$. When using \gls*{qps} to simulate circuits, the variance of the result increases with $\kappa^2$, and we have to compensate for increasing $N_s$ proportionally. This effect is known as sampling overhead. This overhead is multiplicative, increasing exponentially with the number of cut gates $N_c$. Given a large enough number of shots, the outcome of the original circuit is recovered with arbitrary precision. However, noise sources will still introduce a bias in the computation independent of the \gls*{qps}, as noise is a separate quantum channel evolving the state $\rho$. However, quasi-probabilistic simulation techniques have been used to mitigate the effect of noise, again with some sampling cost~\cite{piveteau2022quasiprobability,temme2017error,endo2018practical}, so there is practical overlap between the two techniques. Furthermore, there are some indications that \gls*{qps} can reduce the effect of noise sources by employing smaller circuits~\cite{Ying2023experimental, Singh2023experimental}. Another issue appearing when sampling a \gls*{qps} appears when reconstructing the evolved $\rho$ from the partitions. Due to finite sampling error, finding a distribution with negative terms is possible. To solve this one can apply some post-processing to find the ``most likely'' output state~\cite{Smolin2012mlst, Perlin2021MLFT}.

Finding an efficient \gls*{qpd} of a general circuit $C$, i.e., a \gls*{qpd} with a small $\kappa$, is difficult. If the circuit is known to already have a particular bi-partite structure, one can turn to similar techniques to execute the parts locally, such as Entanglement Forging.~\cite{Eddins2022forging, Failde2023forging}. However, the main direction that has been followed in the literature for circuit cutting was to perform only the \gls*{qpd} of specific regions of the circuit. For instance, simulate only some parts of the circuit that connect regions that are sparsely correlated between them, be it non-local gates or qubit wires.

\paragraph{Circuit cutting techniques: gate-cutting and wire-cutting}

One preliminary work, which was later labeled as circuit cutting (and in particular, wire-cutting), was the \textit{cluster simulation sche\-me} \cite{peng2020simulating}, which decomposes the corresponding tensor network of a given quantum circuit into smaller clusters. Inter-cluster communication is then simulated classically. The authors apply these techniques for Hamiltonian simulation using the \gls*{vqe} \cite{Peruzzo2014}, and suggest using this hybrid variational ansatz for future modular architectures. Later, Mitarai and Fujii~\cite{mitarai2021constructing} introduce the idea of \textit{virtual two-qubit gates}, where the action of the virtual gate is substituted with local operations. This way they only apply \gls*{qps} for the non-local gates we want to get rid of. Given that most \glspl*{qpu} can only execute single- and two-qubit gates, it is more convenient to find an efficient \gls*{qpd} of the particular two-qubit gate and simulate them with local single-qubit gates. The total overhead of the \gls*{qps} then scales as $\mathbb{O}(\kappa^{2 N_c})$ with $N_c$ being the number of virtual gates. Mitarai and Fujii also provide an efficient \gls*{qpd} for a two-qubit gate with $\kappa=3$ at most, from which most common two-qubit gates such as $CNOT$, $CZ$, $RZZ(\theta)$, etc., can be derived. Fig.~\ref{compiler:fig:cutting-schemes} compares the two methods, which can also be used simultaneously in the same circuit. 

The main drawback of circuit cutting is the exponential overhead, so minimizing this quantity is an active research topic. It is important to note however, that this overhead is strictly exponential, and cannot be reduced to a polynomial increase in the number of circuit executions~\cite{Marshall2023bounds}. In~\cite{Brenner2023optimal,piveteau2022circuit}, the minimal sampling overhead to simulate two-qubit gates is derived. Furthermore, ~\cite{piveteau2022circuit} suggests that this overhead can be reduced when jointly cutting multiple gates, using classical communication between the partitions~\cite{piveteau2022circuit}. However, there are recent claims that this classical communication may not be necessary \cite{Schmitt2023multiple}.

Brenner et al.~\cite{Brenner2023optimal} show that cutting an identity gate that transported the state of the qubit before and after the cut is equivalent to a teleportation protocol. As seen in section~\ref{ch6:quantum-teleportation}, to teleport one qubit of data one needs a prepared Bell state and two bits of classical communication. However, gate-cutting of a Bell pair between two qubits (with optimal $\kappa = 3^{N_c}$) is more efficient than cutting a wire, so just by using \gls*{locc} and an ancilla qubit one can optimize the overhead, with an even better scaling for multiple cut wires $\kappa = (2^{N_c+1}-1)$. \gls*{locc} has less demanding hardware requirements than full-on quantum communication with a quantum network. Further studies \cite{Lowe2023fastcutting, Harada2023parallelwirecut, Pednault2023Mar} were able to reduce the ancilla qubits requirement by combining the measure and prepare protocol of wire-cutting, with the idea of classical shadows \cite{Huang2020shadow} and random measurement basis, and \gls*{locc} between the parts. 

A different approach to reduce the sampling overhead in gate-cutting consists of cutting unitaries larger or more complex than two-qubit gates. The search for an optimal \gls*{qps} of a circuit somewhat overlaps with the usual compilation of quantum gates into the native gates of a given quantum computer. For instance, cutting a SWAP gate using \gls*{qps} has a lower sampling overhead ($\kappa = 7^{N_c}$) than first decomposing the SWAP gate into three CNOT gates, and then individually cutting each of them ($\kappa = 3^{3 N_c}$). This can be extended to higher dimension operators, such as multi-controlled CZ gates~\cite{Ufrecht2023multizx}, or even the \gls*{qft}~\cite{Chen2023QFT}. Furthermore, in the case of \gls*{vqa} one can choose variational ansatzes designed with reduced entanglement between parts~\cite{Zhang2022vqereduced, Zhang2023secavqe, Gentinetta2023constrained}, so they are easier to partition.

Other approaches attempt to reduce the number of basis elements of the decompositions to reduce the sampling overhead. Note that, while related in their exponential scaling, the number of subcircuits in a \gls*{qpd} (its 0-norm) is not the same as the sampling overhead (its 1-norm). Reducing the number of subcircuits can help in scheduling and post-processing, but it should be done without increasing the $\kappa$ value. Nagai et al.\ realize this by introducing pre- or post-selection methods for quantum channels~\cite{Nagai2023prepost}, while Chen et al.\ use approximate methods that directly neglect some of the elements~\cite{chen2022approx, Chen2023approx}.

Another separate effort to reduce the overhead comes from minimizing not the \gls*{qpd} of a unitary itself but the amount of quantum communication between machines through smart choice of qubit assignment between machines. For instance, by combining both gate- and wire-cutting techniques, one can find better partitions compared to only using either one of them \cite{Brandhofer2023partitioning}. This is of pivotal interest for \gls*{dqc} in general, not only for circuit cutting, as detailed in section \ref{compiler:subsubsec:circuit-distribution}. The same difficulties and techniques that appear when distributing quantum circuits in a quantum network are mirrored with circuit-cutting protocols. A solution that minimizes the sampling overhead also minimizes the number of Bell pairs in a \gls*{dqc} protocol, and thus, the same compiling tools could be used for both techniques. Furthermore, some \glspl*{sdk}, such as Qiskit or Pennylane, incorporate these techniques in their compilation routines. Moreover, several tools such as CutQC \cite{tang2021cutqc}, ScaleQC \cite{Tang2022scaleqc} or SuperSim \cite{Smith2023supersim} perform the whole circuit cutting pipeline, finding cuts, executing the subcircuit, and reconstructing the state. There is also, as we will delve in section~\ref{compiler:subsec:compilation}, a compiler named Qurzon~\cite{Chatterjee2022-qurzon} which performs all the aforementioned techniques --~in fact, it uses CutQC in combination with other tools.

All in all, finding the optimal \gls*{qps} of a given circuit can become one extra layer of the compilation for quantum circuits, previous to transpilation. Although finding the optimal decomposition is, in general, an NP problem, heuristic methods may find a satisfactory solution. This is another way in which classical computation can further help in reducing the quantum resources of quantum computation.  

\subsubsection{Embarrassingly parallel}
\label{compiler:subsubsec:embarrasingly-parallel}
In the context of quantum computing, the term \textit{embarrassingly parallel} refers to the scenario where a problem can be divided into multiple smaller computations that can be executed independently without the need for direct communication among them. The simplest example of this in the quantum case is the \textit{distribution of shots}, where a quantum algorithm or kernel needs to be executed multiple times without any structural changes --~except for the modification needed to map the circuit to the different \glspl*{qpu}~--. Despite the quantum nature of the tasks involved, this method essentially involves classical parallelism, as was earlier mentioned in the section.

A different approach comes from a distribution of the circuits needed to reconstruct the expectation value of a given observable or to support the optimization protocol. This allows several possibilities:

\begin{enumerate}
    \item \textit{Distribution of terms in an observable}. The distribution of the expectation value terms $\langle O_i \rangle$ of a given observable $\langle O \rangle=\sum \langle O_i \rangle$ is a case of embarrassingly parallelization. An intuitive example is the \gls*{vqe}~\cite{Peruzzo2014}, where the function to minimize is the energy, i.e., the expectation value of a Hamiltonian $\langle H \rangle$. Depending on the specific problem, Hamiltonians can be commonly expressed using fermionic operators in second quantization formalism, as in the case of many systems in condensed matter/chemistry, bosonic operators, or directly in Pauli operators, as in spin Hamiltonians that apply to different problems in physics, route optimization, protein folding \cite{Robert2021}, and scheduling, among others.  In all cases, except the last one, the Hamiltonian has to be mapped to qubit instructions via some encoding techniques \cite{Tilly-VQE-2022, Wang_2023}. After that, it appears as a weighted sum of tensor products of Pauli operators, most commonly known as Pauli strings. Initially, each Pauli string can be individually sent to different \glspl*{qpu}. However, the scaling in the number of Pauli strings for complex problems makes this procedure inefficient. A common practice is to form groups of Pauli strings that will share the same quantum circuit to construct their expectation value. These groups are made of commuting Pauli operators that are determined using some classical routine. The simplest strategy is \textit{qubit-wise commutativity}, where each of the commuting groups built can be measured using a single quantum circuit without difficulties \cite{Tilly-VQE-2022}. An alternative is \textit{general commutativity}, which is more efficient in reducing the number of commuting groups but entails the non-trivial task of finding the appropriate unitaries for the joint measurement of the groups \cite{Tilly-VQE-2022,gokhale2019minimizing}.

    \item \textit{Gradient and Hessianss distribution}. Just like the preparation of a parameterized trial wave function $\ket{\psi(\theta)}$ to our problem, first and second partial derivatives of the state $\ket{\psi(\theta)}$ can be analyzed with a quantum computer~\cite{PhysRevA.99.032331,  QNG, viqueira2023density}. In many cases, the quantum circuits that arise from the partial derivatives can be expressed as a linear combination of circuits that use the same structure of the original circuit to prepare $\ket{\psi(\theta)}$, with a shift in their parameters, which is known as parameter shift rule~\cite{Wierichs2022generalparameter}.

    \item \textit{Distribution in a gradient-free optimization}. That is a particular case of distribution that sources from the usage of gradient-free optimizers such as evolutionary optimizers. These optimizers overcome the need to compute gradients at the cost of using several individuals/particles that interact in a certain way to modify their parameters or generate other candidates. That is the case, for example, of Differential Evolution and the Particle Swarm Optimization algorithms~\cite{Robert2021, Faílde2023, Alvarez-Alvarado2021}. Each individual is a different set of parameters that can be executed in parallel using the same quantum circuit structure. One of the possible benefits of the previously mentioned optimizers is that they can mitigate problems in the optimization landscape  \cite{Faílde2023,Anschuetz2022}. However, this would come at the cost of increasing drastically the number of circuit executions.
    
    \item \textit{Distribution of data}. As in the case of classical Machine Learning, another possibility is to distribute the data or the model during the training. For example,~\cite{Du2022} proposes a tool for distributing training of Quantum Machine Learning models that can also be used for \glspl*{vqe}. A federated approach has also been proposed \cite{Chen2021}.
\end{enumerate}

There are some packages that permit the distribution of these kinds of jobs among several \glspl*{qpu} \cite{Du2022,Stein2022}, based on a master-worker architecture. These packages must cope with additional issues not seen in classical Machine Learning distributed learning, such as the different architectures of the \glspl*{qpu} (different gate sets, different topology, or different timing for execution), the noise of each single QPU and the possible drift of these errors with the time, for counting some of the current challenges. Additionally, these techniques can also be used when circuit cutting is applied.

Another paradigm that can be considered in this context is \textit{multi-programming} of quantum computers. The segmentation of a QPU, better known as multi-pro\-gra\-mming in quantum computing, can maximize the hardware throughput --~the number of used qubits divided by the total number of qubits~-- and reduce the runtime. The pioneering work for multi-programming by Das et al.\ \cite{das2019case} advocated for the use of multi-programming to enhance the utilization and throughput of NISQ computers, wherein the qubits are employed to execute multiple workloads concurrently. It also presented various techniques that will be further elucidated in future sections and with which the hardware throughput of IBM-Q16 was improved. Other works introduce enhancements like selecting the appropriate number of circuits to execute, qubit mapping, device benchmarking, cross\-talk\footnote{Crosstalk is an unwanted coupling between qubits. It is one of the noise sources in NISQ devices and can condition the hardware throughput.} characterization, or even vulnerability analysis \cite{ash2020analysis, liu2021qucloud, ohkura2021crosstalk,saki2021qubit, niu2022multi, niu2022parallel,niu2023enabling, niu2023multi, liu2024qucloud+, saki2021shuttle}. Again, we will describe some of these works when talking about the compilation process.

Another paradigm that may be interesting to delve into is \textit{quantum offloading}. As mentioned in the introduction, \glspl*{qpu} is intended to be seamlessly integrated into classical \gls*{hpc} infrastructures, working along other hardware accelerators. This way of distributing the workload allows concurrent computations of classical and quantum tasks, letting CPUs proceed with calculations while \glspl*{qpu} accelerate specific processes in which the so-called \textit{quantum advantage} takes part.

A profound quantum offloading analysis diverges from this work's main scope, but some relevant works can be outlined. For instance, the \gls*{xacc} is a system-level software infrastructure for quan\-tum-cla\-ssical computing that promotes a service-oriented architecture to expose interfaces for core quantum programming, compilation, and execution tasks \cite{McCaskey2019XACC:Computing}. Strongly related is QCOR, a language extension specification of C++ that enables single-source quantum-clas\-si\-cal programming and that employs \gls*{xacc} as a base \cite{mccaskey2021-qcor}. Another work leveraged the OpenMP API to target quantum devices, which provides an easy-to-use and efficient interface for HPC applications to utilize quantum computing resources \cite{lee2023quantum}. Similar to this were the efforts made to add \glspl*{qpu} to the OpenCL ecosystem of execution \cite{vazquez2024qpu}. Even the NVIDIA company has developed the CUDA Quantum Platform for hybrid quantum-classical computation, enabling the aforementioned integration and programming of \glspl*{qpu} along with other accelerators.

\subsection{Compilation}
\label{compiler:subsec:compilation}
After resolving the distribution challenge, it is essential to explore the compilation process thoroughly. We will adhere to a structure akin to the classical approach, which involves an analysis phase, an intermediate representation referred to as \gls*{qir}, and a synthesis phase. This framework will aid in comprehending the compilation process for \gls*{dqc} and underscore the disparities between classical and quantum computing in terms of compilation.

\subsubsection{Analysis phase}
\label{compiler:subsubsec:analysis-phase}
The analysis phase in the distributed and monolithic quantum compilation is quite similar, with the additional challenge in the distributed case of limited literature and software development compared to the monolithic counterpart. In the monolithic scenario, the underutilization of standalone languages is not because they do not exist; rather, options like Scaffold~\cite{abhari2012-scaffold}, Q\#~\cite{Svore2018-QSharp}, isQ~\cite{guo2023-isQ}, Q$|SI\rangle$~\cite{liu2018-qsi}, among others, are available. However, they are less favored due to the need for users to understand and adapt to these languages. In contrast, libraries like Qiskit~\cite{Qiskit2023}, Cirq~\cite{Developers2023Cirq} and Qulacs~\cite{suzuki2021-qulacs}, built on well-known classical languages such as Python (Qiskit and Cirq) and C++ (Qulacs), are more widely adopted. This situation is even more pronounced in the distributed case because there is a shortage of standalone languages specifically designed for distributed purposes. Consequently, the previously mentioned quantum monolithic libraries are often repurposed to simulate the distributed structure.

This is the case for \gls*{qmpi} \cite{Haner2021DistributedQMPI}, which represents an extension of the \gls*{mpi} protocol for distributed quantum systems. We refer to this as a formal approach due to the absence of a usable library that allows for actual or simulated \gls*{dqc}. However, a reference implementation for \gls*{qmpi} has recently been published \cite{Shi2023}, although none of the code is available for use, neither in open source nor as a binary, to the best of our knowledge. 

The aim of \gls*{qmpi} is, obviously, to add quantum functionalities to an already widely used specification such as \gls*{mpi}. For this purpose, it defines two types of nodes: classical and quantum. The only difference between them is that classical nodes cannot be the target of quantum directives, whereas quantum nodes can manage both quantum and classical calls. The core of this difference lies in the inherent distinction between classical datatypes and quantum datatypes --~bits and qubits~-- along with the inclusion of EPR pairs, a crucial element for the development of quantum communication protocols, as shown in section~\ref{sec:physical}. Other than that, although \gls*{mpi} is much more advanced than \gls*{qmpi}, as expected, the communication modes supported by the latter are the same: point-to-point communication and collective operations. Moreover, they define a simple performance model called SENDQ. It is worth mentioning that, contrary to almost all literature on \gls*{dqc}, they anticipate a relatively low logical clock speed for quantum computers due to the overhead introduced by the quantum error correction. Consequently, they do not expect classical communication to significantly affect performance, choosing to ignore classical communication in the SENDQ model. This approach contrasts significantly with all the circuit distribution methodologies discussed in section \ref{compiler:subsubsec:circuit-distribution}, where the focus is primarily on minimizing the number of teledata and telegates, considered the main bottleneck of quantum distribution --~as was mention early in that section. Their SENDQ model is closely associated with the \gls*{nisq} era and may not be sustainable when transitioning to the fault-tolerant era.

Anyway, as it is explained in Wakizaka~\cite{Wakizaka2023}, there is a need to develop a proper quantum programming language that takes consideration of a distributed structure and extracts profit from that structure via advanced distributed computational techniques, just as it happens in classical computation. 

\subsubsection{Distributed quantum Intermediate Representation}
\begin{figure}
    \centering
    \includegraphics[width=0.8\linewidth]{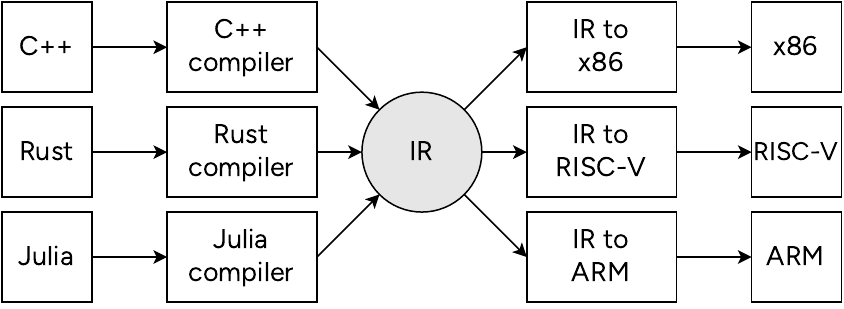}
    \caption{The significance of intermediate representation in the compilation process - Facilitating decoupling between high-level and machine code.}
    \label{compiler:fig:decoupling}
\end{figure}

The compilation process is complex, therefore \aclp*{ir} (\acs*{ir}) were introduced to establish a break in the compiler in order to obtain modularity and decoupling~\cite{Chow2013}. An \gls*{ir} allows to intermediate between the front-end and the back-end, improving the efficiency of compiler development and allowing abstract optimizations to the target machine. Fig.~\ref{compiler:fig:decoupling} shows the use of \acsp*{ir} as a break in the compilation process to facilitate compiler development so that programs are implemented for abstract machine code such as an \acs*{ir}.

An important feature of \acsp*{ir} is that they have to be able to represent the operations of different high-level languages to be implemented in different machine codes. Therefore, with the evolution of quantum computing, it is necessary to extend classical \acsp*{ir} (or create new ones) to include quantum instructions. This process has been evolving in recent years, where the number of quantum IRs has grown considerably~\cite{McCaskey2021, Lubinski2022, Peduri2022, IttahDavid2022QIRO:Optimization, Hietala2019, luo2020-yao}.

For \gls*{dqc}, specialized \gls*{ir} are needed to allow the use of classical and quantum communication instructions between different \acsp*{pu}. This objective is what InQuIR~\cite{Nishio2023-inquir}, an \gls*{ir} specialized in \gls*{dqc}, aims to solve.

To exemplify the operation of this \gls*{ir}, we use the circuit shown in Fig.\   \ref{subfig:circuits_tele:telegate}, which implements a CNOT remote gate between two separate nodes, but connected through a Bell pair $|\Phi^+\rangle$. Fig.~\ref{compiler:code:openqasm-example} shows the OpenQASM code to implement this, which does not consider communication directives. The compilation of OpenQASM to InQuIR produces the code shown in Fig.\ \ref{compiler:code:inquir-node0-example} for node 0 and Fig.\ \ref{compiler:code:inquir-node1-example} for node 1. InQuIR automatically adds the necessary directives to do the remote operation using the telegate technique.

\begin{figure}[t]
    \centering
    \begin{subfigure}[b]{0.65\linewidth}
        \centering
        \begin{lstlisting}[language=llvm,frame=single,breaklines=true,gobble=7,basicstyle=\scriptsize\ttfamily]
        OPENQASM 2.0;
        qreg q[2];
        h q[0];
        cx q[0],q[1];\end{lstlisting}
        \caption{OpenQASM 2.0 code for the creation of an EPR pair.}
        \label{compiler:code:openqasm-example}
    \end{subfigure}
    
    \begin{subfigure}[b]{0.65\linewidth}
        \centering
        \begin{lstlisting}[language=llvm,frame=single,breaklines=true,gobble=7,basicstyle=\scriptsize\ttfamily]
        0 {
          world = open[0,1];
          q0 = init();
          _cq0 = genEnt[1](l0);
          CX q0 _cq0;
          _m0 = measure _cq0;
          free _cq0;
          send[1](world, l1:_m0);
          recv(world, l1_2:_m1);
          Z[_m1] q0;
        }\end{lstlisting}
        \caption{InQuIR code for node 0 (qubit 0).}
        \label{compiler:code:inquir-node0-example}
    \end{subfigure}

    \begin{subfigure}[b]{0.65\linewidth}
        \centering
        \begin{lstlisting}[language=llvm,frame=single,breaklines=true,gobble=7,basicstyle=\scriptsize\ttfamily]
        1 {
          world = open[0,1];
          q1 = init();
          _cq1 = genEnt[0](l0);
          CX _cq1 q1;
          H _cq1;
          _m2 = measure _cq1;
          free _cq1;
          send[0](world, l1_2:_m2);
          recv(world, l1:_m3);
          X[_m3] q1;
        }\end{lstlisting}
        \caption{InQuIR code for node 1 (qubit 1).}
        \label{compiler:code:inquir-node1-example}
    \end{subfigure}
    \caption{InQuIR representation of the creation of an EPR pair using remote gates .}
    \label{compiler:fig:inquir-creation-epr}
\end{figure}

The IR code extends the basic quantum operations to a distributed setting, where quantum communication and entanglement generation across different nodes (0 and 1) are involved. Lines 2 to 4 in both figures \ref{compiler:code:inquir-node0-example} and \ref{compiler:code:inquir-node1-example} correspond to the initialization of the communication channel between both nodes, the initialization of the local qubits, and the generation of the EPR pair, respectively. Lines 5--6 in \ref{compiler:code:inquir-node0-example} and 5--7 in \ref{compiler:code:inquir-node1-example} correspond to the gates and measurements.  The measurement results are transferred between the two nodes by \texttt{send}/\texttt{recv} operations and used in the conditional gates.

\subsubsection{Synthesis phase}
\label{compiler:subsubsec:synthesis-phase}
In classical compilation, this corresponds to the lowest level of abstraction. In quantum compilation, nevertheless, it is difficult to associate each of the quantum compilation stages to a different level of abstraction because there are almost no abstraction layers in the quantum programming ecosystem. But, as a parallelism to classical compiling, we can associate this stage to the \gls*{qasm}. There are a lot of different versions, such as OpenQASM~\cite{Cross2017-OpenQASM}, cQASM~\cite{Khammassi2018-cQASM}, eQASM~\cite{fu2019-eQASM} and f-QASM~\cite{Svore2006-fQASM}. But, to the best of our knowledge, only  NetQASM \cite{Dahlberg2021NetQASMInternet} takes into account an underlying distributed structure.

In \cite{Dahlberg2021NetQASMInternet}, Dahlberg et al.\ introduced an abstract model featuring a \gls*{qnpu} for end-nodes in a \gls*{qn}. NetQASM is proposed as an \gls*{isa} designed to execute arbitrary programs on end nodes equipped with the \gls*{qnpu}. So, NetQASM can be seen as a low-level, assembly-like language tailored for the quantum segments of quantum network program code. It specifies the interaction between the \gls*{qnpu} and executes \gls*{qn} code, a functionality not available in other \gls*{qasm} languages. The language is designed to be extensible, with a core set of instructions for classical control and memory operations, and a set of quantum-specific instructions grouped into ``flavors''. A ``vanilla'' flavor is introduced for universal, platform-independent quantum gates, enabling platform-independent quantum network program descriptions, with the possibility of developing platform-specific flavors for optimized quantum operations on specific hardware, such as \gls*{nv} hardware for quantum network end-nodes recalling from section \ref{sec:physical}.

It is also worth mentioning the work of Ying and Feng \cite{ying2009algebraic}. They developed an algebraic language for formally specifying quantum circuits in \gls*{dqc} that aims to represent circuits conveniently and compactly, akin to how Boolean expressions are used for classical circuits.

Delving now into the synthesis phase of quantum compilation, this phase can be broadly divided into three main components: \textit{optimization}, \textit{verification}, and \textit{qubit mapping}. 
Circuit optimization involves reducing circuit complexity based on a specific metric, which often measures quantum computations' efficiency and error susceptibility. This is especially critical in the current \gls*{nisq} era, where quantum hardware has significant limitations.  Circuit optimization is a really complex field of study, especially in the monolithic case. Of the other two stages, circuit verification is responsible for checking whether the quantum circuit performs the correct computations. In the classical world, this responsibility does not usually fall on the compiler, but on the debugger. On the other hand, qubit mapping focuses on how the logical qubits of a quantum algorithm are mapped to the physical qubits of a quantum processor or, specifically in \gls*{dqc}, a set of interconnected processors. 

\paragraph{Optimization}
The optimization phase in monolithic quantum computing encompasses a broad range of techniques aimed at minimizing various metrics, such as the number of 2-qubit gates, the circuit depth, etc. In \gls*{dqc}, we encounter similar optimization challenges as in the monolithic case, but with the added complexity of distributing or cutting the circuits. On the contrary, if the distribution technique performed is embarrassingly parallel, the optimization phase is, naturally, equivalent to the monolithic one, excepting the case of multi-programming where optimizations are subtle and tend to be related with crosstalk and fidelity \cite{liu2021qucloud, liu2024qucloud+}.

Delving into circuit distribution, we have discussed in section \ref{compiler:subsubsec:circuit-distribution} the circuit distribution methods and efforts made to partition the circuit optimally before performing local mapping. In essence, optimization in this case mirrors that of the monolithic case, but with the additional consideration of the partitioning problem, which is intricately linked to qubit mapping. Indeed, the close relationship between qubit mapping and circuit optimization is not surprising, even in the monolithic case. It is logical because an efficient mapping of qubits directly impacts circuit performance, much like how effective register management optimizes classical computing tasks. However, although we are only adding one more constraint with the circuit distribution, it is of vital importance since the teleport and telegate costs are significantly higher than those of local 2-qubit gates. As previously discussed in Section \ref{compiler:subsubsec:circuit-distribution}, this serves as justification for why circuit partitioning methods consistently aim to minimize the utilization of these remote protocols. Qiu and Chen \cite{Qiu2022} realize an interesting analysis of this topic, where the quantum cost figure of merit is employed. The quantum cost of a circuit is calculated by summing the cost of each gate present in the circuit. Any gate can be broken down into several basic gates, each with a unit cost, irrespective of their internal complexity. Using this definition of cost they showed the expensiveness of quantum teleportation and dense coding. However, circling back to the main topic, while we have extensively covered and will further discuss partitioning in the qubit mapping section, we have deliberately chosen not to get deeply into the intricate domain of monolithic quantum optimizations, as it exceeds the scope of this work.

Regarding circuit cutting, optimizations aim at reducing the sampling overhead, or the number of subcircuits. Although both quantities are related in that both increase exponentially with the number of shots, in general, they do not need to scale the same way. The most important of the two is the sampling overhead. Still, a reduction of the number of subcircuits (without an increase in the sampling overhead) can also help in the scheduling and post-processing part of the computation. Some works reduce the sampling overhead by including \gls*{locc}, either when jointly cutting several gates \cite{Piveteau2022CircuitCommunication}, or in smart prepare-and-measure protocols in wire-cutting \cite{Lowe2023fastcutting, Harada2023parallelwirecut, Pednault2023Mar}. Other works attempt to cut larger unitaries \cite{Ufrecht2023multizx} or constrain the overhead using parameterized gates \cite{Gentinetta2023constrained}. Regarding the number of subcircuits, they can be reduced using pre- or post-selection methods \cite{Nagai2023prepost}, and some of them can be neglected in approximated methods without incurring in large errors \cite{chen2022approx,Chen2023approx}.  

\paragraph{Verification}
Verification of quantum programs is a significant part of quantum compiling. Unlike in the classical world, where developers rely on debuggers to identify and fix errors, debugging quantum programs is inherently difficult due to the destructive nature of measurement. Once a quantum state is measured, it collapses irreversibly, making it impossible to observe the state at different time steps without altering it. Therefore, the verification of quantum programs becomes crucial for ensuring the correct functionality of a quantum circuit. It is essential to incorporate this verification step as a phase in the synthesis stage of compilation. This ensures that the circuit is checked immediately before execution and after optimizations have been applied, to confirm that those optimizations have not altered the functionality of the quantum circuit. In the monolithic realm, several approaches have been made combining optimization and verification in what is usually referred to as \textit{verified optimization} \cite{Fagan2019, Xu2022-Quartz, Hietala2019}.

One way of verifying quantum programs is using quantum process algebras, which are derivations of the classical process algebras. Process algebras, also known as process calculi, are mathematically rigorous languages with well-defined semantics that allow the description and verification of properties of concurrent communicating systems, including, in this case, quantum systems.

There are some examples of these types of formal methods. For instance, \gls*{eqpalg} \cite{Haider2020-eQPAlg}, which extends \gls*{qpalg} \cite{Lalire2004-QPAlg}. More specifically, \gls*{qpalg} provides a homogeneous style for formal descriptions of concurrent and distributed computations, encompassing both quantum and classical components. As authors claim, \gls*{qpalg} introduces quantum variables, operations on these variables --~unitary operators and measurement observa\-bles~-- as well as different forms of communication involving the quantum realm. The operational semantics ensure that these quantum objects, operations, and communications adhere to the postulates of quantum mechanics. Regarding \gls*{eqpalg}, it extends the previous formal specification to accommodate the concept of formally specifying the quantum teleportation protocol, which has been shown in this work to be a key part of the quantum distribution model. The relationship between quantum process algebras and the algebraic language defined in the aforementioned work by Ying and Feng \cite{ying2009algebraic} can be compared to that between classical process algebras and Boolean algebra. In broad terms, quantum process algebras are well-suited for high-level formal specification of \gls*{dqc}, while the language Ying and Feng paper is mainly intended to describe low-level circuit implementation.

Regarding the verification of distributed quantum programs, the work of Feng et al.\ \cite{Feng2022} introduced a distributed programming language designed for formalizing and verifying distributed quantum systems. They presented a Hoare-style\footnote{Hoare logic is indeed a formal system equipped with a set of logical rules used for rigorous reasoning about the correctness of computer programs \cite{Hoare1969}.} logic that is both sound and complete, aiding in the analysis and verification of quantum programs, including quantum teleportation and CNOT gates. Talking specifically about distributed quantum protocols, Wang's work \cite{Wang2021verification} profoundly delves into the verification of several distributed quantum protocols such as the BB84 protocol \cite{Bennett2014}.

\paragraph{Qubit mapping}
When it comes to classic computing, register allocation is about finding the best way to use the limited number of registers available to store variables \cite{Wall1986}. In the field of quantum computing, qubit mapping can be compared to register allocation in classical computing. This process involves finding an optimal mapping of logical qubits to physical qubits in a quantum device, taking into account the device's connectivity and other constraints. It is important to note the growth in complexity of this process as it moves from classical to quantum compilation. In the realm of quantum compilation, it is not only the use of the qubit's value that must be evaluated --~meaning if it is thought to be a communication qubit or a computing qubit. Other factors, such as the error associated with the specific qubit and its interconnection with the remaining qubits, assume significance in the decision-making process. Qubit mapping is an NP-hard problem \cite{Ito2023}. Therefore, exact algorithms are only computable for a reduced number of qubits, making it necessary to use techniques that are able to obtain an optimal solution even if it is not the best one. Additionally, the quantum mapping process can be separated into three processes:

\begin{itemize}
    \item\textit{Gate decomposition:} Refers to the stage in which gates composing the circuit are transformed into a series of native gates implementable in the actual quantum processor. This is one of the aforementioned device's constraints that have been taken into account. 
    \item\textit{Quantum allocation:} Refers to the process of physically assigning specific logical qubits in a quantum processor. For a correct qubit allocation, in most cases, it is necessary to add additional SWAP gates to move the qubit information~\cite{Paler2018}.
    \item\textit{Quantum routing:} Refers to the task of finding efficient paths for communication between qubits in a quantum processor. This is important when mapping gates of two logic qubits that are not interconnected to maximize efficiency~\cite{Bandic2023, Cowtan2019}. For a thorough analysis of the qubit routing problem, one can check the review on the subject by Barnes~\cite{Barnes2023}.
\end{itemize}

It is also common to consider a fourth stage called \textit{gate scheduling}, which tries to leverage parallelism while respecting dependencies and quantum hardware constraints. Fig.~\ref{compiler:fig:qubit-routing-example} shows a specific qubit routing problem in which a qubit allocation has already been performed. Figures \ref{compiler:subfig:interconnection1} and \ref{compiler:subfig:interconnection2} show a ring type qubit interconnection network --~only communication with adjacent qubits~-- and a one-way architecture --~received from one neighbor and sent to another~--, respectively. With these network architectures, the logic circuit shown in Fig.\ \ref{compiler:subfig:logic-circuit} will be transformed into an equivalent circuit that meets the connectivity constraints. Fig.~\ref{compiler:subfig:interactions} shows that the constraints are being violated by performing a CNOT gate between $q_{3}$ and $q_{1}$. A solution to this constraint is shown in Fig.\ \ref{compiler:subfig:sol-interconnection1} for the ring network architecture. Here, a swap gate is used to interchange $q_{1}$ and $q_2$, which allows the CNOT operation to be performed between $q_3$ and $q_2$ and, finally, a new swap recovers the $q_2$ state. A CNOT gate cannot be performed in the direction $q_{3}$ to $q_{1}$ for the one-way network architecture. Therefore, it is necessary to use a mechanism as shown in Fig.\ \ref{compiler:subfig:sol-interconnection2} to reverse the gate order.

\begin{figure*}[t]
    \centering
    \begin{subfigure}[b]{0.48\linewidth}
    \centering
        \begin{tikzpicture}[scale=0.75,every node/.style={transform shape}]
            \foreach \i in {1,...,8}
                \node[circle, draw] (q\i) at (-360/8*\i-360/8-180:1.5) {\(q_{\i}\)};
            \foreach \i in {1,...,7}
                \draw (q\i) -- (q\the\numexpr\i+1\relax);
            \draw (q8) -- (q1);
        \end{tikzpicture}
    \caption{Physical ring interconnection network of the quantum chip.}
    \label{compiler:subfig:interconnection1}
    \end{subfigure}
    \hspace{0.5cm}
    \begin{subfigure}[b]{0.48\linewidth}
    \centering
        \begin{tikzpicture}[scale=0.75,every node/.style={transform shape}]
            \foreach \i in {1,...,8}
                \node[circle, draw] (q\i) at (-360/8*\i-360/8-180:1.5) {\(q_{\i}\)};
            \foreach \i in {1,...,7}
                \draw[->] (q\i) -- (q\the\numexpr\i+1\relax);
            \draw[->] (q8) -- (q1);
        \end{tikzpicture}
        \caption{Physical one-way interconnection network of the quantum chip.}
        \label{compiler:subfig:interconnection2}
    \end{subfigure}

    \hfill
    
    \begin{subfigure}[b]{0.48\linewidth}
        \centering
        \begin{adjustbox}{scale=0.75}
        \begin{quantikz}
            \lstick{$q_1$} & \gate{H} & \ctrl{1} & \targ{}\gategroup[wires=3, steps=1, style={dashed, rounded corners, fill=red!10, inner sep=2pt}, background]{CNOT not connected} & \qw & \qw & \qw \\
            \lstick{$q_2$} & \qw & \targ{} & \qw & \qw & \qw & \qw \\
            \lstick{$q_3$} & \gate{H} & \ctrl{1} & \ctrl{-2} & \qw & \qw & \qw \\
            \lstick{$q_4$} & \qw & \targ{} & \qw & \qw & \qw & \qw
        \end{quantikz}
        \end{adjustbox}
        \caption{Logic circuit for which $q_{i}$ already indicates the physical qubit $i$.}
        \label{compiler:subfig:logic-circuit}
    \end{subfigure}
    \hspace{0.5cm}
    \begin{subfigure}[b]{0.48\linewidth}
    \centering
        \begin{tikzpicture}[scale=0.75,every node/.style={transform shape}]
        \foreach \i in {1,...,4}
                \node[circle, draw] (q\i) at (-360/8*\i-360/8-180:1.5) {\(q_{\i}\)};
            \draw[->] (q1) -- (q2);
            \draw[->] (q3) -- (q4);
            \draw[->, red, dashed] (q3) -- (q1);
        \end{tikzpicture}
        \vspace{0.5cm}
        \caption{Graph representing the interactions between the physical qubits of the logic circuit.}
        \label{compiler:subfig:interactions}
    \end{subfigure}

    \hfill
    
    \begin{subfigure}[b]{0.48\linewidth}
        \centering
        \begin{adjustbox}{scale=0.75}
        \begin{quantikz}
            \lstick{$q_1$} & \gate{H} & \ctrl{1} & \swap{1}\gategroup[wires=3, steps = 3, style={dashed, rounded corners, fill=green!10, inner sep=2pt}, background]{CNOT connected} & \qw  & \swap{1} \\
            \lstick{$q_2$} & \qw & \targ{} & \targX{} & \targ{} & \targX{} \\
            \lstick{$q_3$} & \gate{H} & \ctrl{1} & \qw & \ctrl{-1} & \qw \\
            \lstick{$q_4$} & \qw & \targ{} & \qw & \qw & \qw
        \end{quantikz}
        \end{adjustbox}
        \caption{Circuit transformation for the physical ring network.}
        \label{compiler:subfig:sol-interconnection1}
    \end{subfigure}
    \hspace{0.5cm}
    \begin{subfigure}[b]{0.48\linewidth}
        \centering
        \begin{adjustbox}{scale=0.75}
        \begin{quantikz}
            \lstick{$q_1$} & \gate{H} & \ctrl{1} & \swap{1}\gategroup[wires=3, steps=5, style={dashed, rounded corners, fill=green!10, inner sep=2pt}, background]{CNOT connected} & \qw  & \qw & \qw & \swap{1} \\
            \lstick{$q_2$} & \qw & \targ{} & \targX{} & \gate{H} & \ctrl{1} & \gate{H} & \targX{} \\
            \lstick{$q_3$} & \gate{H} & \ctrl{1} & \qw & \gate{H} & \targ{} & \gate{H} & \qw \\
            \lstick{$q_4$} & \qw & \targ{} & \qw & \qw & \qw & \qw & \qw
        \end{quantikz}
        \end{adjustbox}
        \caption{Circuit transformation for the physical one-way network.}
         \label{compiler:subfig:sol-interconnection2}
    \end{subfigure}
    
    \caption{Example of the transformation of a logic circuit to match two physical network architectures for interconnecting qubits: {(\subref{compiler:subfig:interconnection1},\subref{compiler:subfig:interconnection2})} two examples of graphs indicating the connections between the physical qubits on the chip, a ring connection on the one hand and a one-way connection on the other; {(\subref{compiler:subfig:logic-circuit},\subref{compiler:subfig:interactions})} example of a logic circuit with a CNOT gate between two qubits that are not connected and the interaction graph between the qubits generated by the circuit; {(\subref{compiler:subfig:sol-interconnection1},\subref{compiler:subfig:sol-interconnection2})}  transformations applied to obtain an equivalent circuit complying with the interconnection network constraints of each example (ring and one-way).}
    \label{compiler:fig:qubit-routing-example}
\end{figure*}

Regarding \gls*{dqc}, it is essential to distinguish between distribution methods that require partitioning and those that do not. In the former case, where partitioning is necessary, the qubit mapping problem aligns with the classical problem. Still, it includes the additional challenge of optimizing circuit partitioning to minimize communication, as detailed in section \ref{compiler:subsubsec:circuit-distribution}, where we already mentioned how linked are those methods with this stage of compilation. Indeed, it may seem repetitive, but it is crucial to emphasize the significant impact of the circuit partitioning method across all stages of distributed quantum compilation.

Nevertheless, a few works that have not been mentioned in that section are of interest. The first one is the work of Mao et al.\ \cite{Mao2023-qubit-allocation}, which baptizes the problem as \gls*{qa-dqc}, proves the NP-hardness of it and proposes two algorithms to deal with it: a heuristic local search algorithm and a \gls*{mhsa} algorithm. In the latter, they combine the local search algorithm and a simulated annealing meta-heuristic algorithm, along with extensive simulations to evaluate it. The second work is also carried out by Mao et al.\ \cite{Mao2023} that proposes a probability-aware qubit-to-processor mapping model, which incorporates communication overhead between processor pairs determined through probabilistic analyses based on link entanglement generation rates. Additionally, they introduced a multi-flow routing protocol to enhance overall entanglement rates. Subsequently, they employed a multistage hybrid simulated annealing algorithm, which is reminiscent of the previous one, to minimize total communication overhead. As we have already mentioned, extensive simulations are conducted to demonstrate the effectiveness of these solutions across various system settings. The third work of interest in this line is the one developed by Nakai \cite{Nakai2022}, which deeply develops the qubit allocation problem for \gls*{dqc} along with a formal definition of the problem as an optimization problem similar to how we have defined the partitioning one. And, finally, the last work is developed by Chen et al.\ \cite{chen2023routing} where they focus on the step following the circuit partitioning, i.e., the qubit routing stage. Specifically, they focused on investigating the influence of the quantum state transmission direction during the execution of global gates on the number of transmissions and subsequent routing. It utilizes a heuristic algorithm, called \gls*{gagdo}, to ascertain the optimal transmission direction for all global gates in the circuit, with the goal of minimizing the overall cost of the executable circuit generated in the distributed architecture model.

Also, two works have been developed to characterize the inter-core qubit traffic in which some benchmarks arise in order to analyze mapping performance \cite{rodrigo2022characterizing, rached2023characterizing}. They employed the OpenQL compiler \cite{khammassi2021openql}, which is not a distributed compiler \textit{per se}, but allows the embedding of a modified version of the Qmap mapper \cite{lao2021timing}. In particular, for this case, they extended it to the multi-core case employing the proposal by Baker et al.\ \cite{Baker2020Time-slicedArchitectures},i.e., the \gls*{fgp}-\gls*{roee} algorithm, already explained in section \ref{compiler:subsubsec:circuit-distribution}.

Now, in cases of embarrassingly parallel distribution, where partitioning is not required, the qubit mapping problem mirrors that of the monolithic case, with the added complexity of needing to perform mapping for each \gls*{qpu}. This complexity arises from the potential differences in architectures among the \glspl*{qpu} contained in the distributed scheme. There is just one case in the embarrassingly parallel scenario where qubit mapping differs from the monolithic case: the multi-programming scenario. This paradigm of quantum execution, which involves segmenting the \gls*{qpu}, imposes a series of constraints on the qubit mapping problem. One of the first approaches was the already mentioned work by Das et al.\ \cite{das2019case}. Three techniques were developed in this work:

\begin{enumerate}
    \item \gls*{frp} algorithms, developed to partition qubit resources into multiple groups fairly, while avoiding qubits or links with excessively high error rates.
    \item \gls*{dis} policy, devised to mitigate interference from measurement operations of one program on the gate operations of co-running programs.
    \item \gls*{amp} design, proposed to monitor reliability impact at runtime and revert the system to isolated execution mode if the impact is high.
\end{enumerate}

Different techniques were developed under the QuCloud fra\-mework by Liu and Dou \cite{liu2021qucloud}. In this work, they developed, also, three approaches:

\begin{enumerate}
    \item They utilized community detection techniques to partition physical qubits among concurrent quantum programs, mitigating resource waste. They even proposed a new technology based on these techniques called \gls*{cdap}.
    \item They designed the X-SWAP scheme, which enables inter-program SWAPs and gives priority to SWAPs linked with critical gates to minimize SWAP overheads.
    \item They introduced a compilation task scheduler that prioritizes the compilation and execution of concurrent quantum programs based on estimated fidelity for optimal performance. 
\end{enumerate}

This was further extended in a subsequent work by the same authors under the QuCloud+ framework \cite{liu2024qucloud+}, in which they tried to take into consideration the crosstalk effect on real-world applications.

\subsubsection{Available compilers}
\label{compiler:subsubsec:compilers}
Not many full-stack tools or compilers are designed considering a distributed quantum scheme as a base. In fact, to the best of our knowledge, there is no compiler for \gls*{dqc} available for use, just conceptual designs and prototypes. These conceptual quantum compilers can be classified depending on which type of distribution they use from the ones described in section\ \ref{compiler:subsec:compilation}, i.e., usual circuit distribution, circuit cutting, and embarrassing parallelism.

\paragraph{Compilers for circuit distribution}
Ferrari et al.\ \cite{Ferrari2021} designed a distributed quantum compiler that focuses on the minimization of the depth of the circuit and, for this matter, two different techniques are tested: \textit{data-qubit-swapping-based strategy} and \textit{entanglement-swapping-based strategy}. They compared the performance of the partitioning --~and, hence, of the distribu\-tion --~of these two strategies with the already analyzed work by Martinez and Heunen \cite{Andres-Martinez2019AutomatedPartitioning}. Also, Ferrari et al.\ \cite{Ferrari2023} designed a versatile modular quantum compilation framework for \gls*{dqc}, which considers both network and device constraints and characteristics. For qubit assignment, they employed METIS’s multilevel $k$-way partitioning. Moreover, for gate scheduling, they implemented an algorithm to minimize the consumed \gls*{epr} pairs and a local routing algorithm that scans the circuit and, for every gate that involves qubits not directly connected on their specific QPU, it computes the shortest sequence of necessary SWAP gates. The experimental evaluation of a quantum compiler based on this framework was demonstrated, using circuits of interest such as \gls*{vqe}, \gls*{qft}, and graph state preparation, characterized by varying widths --~ ranging from 0 up to 600 qubits. 

Cuomo et al. \cite{Cuomo2023} model the compilation problem using an Integer Linear Programming formulation inspired by the extensive theory on dynamic network problems. They define the problem as a generalization of the quickest multi-commodity flow, enabling optimization using techniques from the literature, such as a time-expanded representation of the distributed architecture. This approach, which also incorporates quasi-parallelism\footnote{The authors define quasi-parallelism as a relaxed version of parallelism based on grouping logically sequenced gates within the same time step.}, allows for more efficient circuit operation and broader solution exploration. The work is modular, enabling adaptation to circuits with varying degrees of operation commutativity and leveraging existing network flow literature. The study aims to refine compiler efficiency and performance through an in-depth analysis of quantum circuits and focus on normal forms. Testing on square and hexagonal lattice topologies showed that square lattices offer superior performance, attributed to their favorable edges-to-nodes ratio, indicating promising avenues for future quantum computing advancements.

\paragraph{Compilers for circuit cutting}
As for now, the only quantum compiler considering the circuit-cutting strategy, as was explained in section \ref{compiler:subsubsec:circuit-cutting} is \textit{Qurzon}~\cite{Chatterjee2022-qurzon}. For the first part of the compilation, an algorithm responsible for cutting the circuit into optimal parts is employed, called CutQC \cite{tang2021cutqc}. After the circuit is cut into several pieces, a scheduling algorithm is responsible for the execution of each of the pieces in the available quantum devices. This problem is nothing more than a classic problem of scheduling jobs, well known in the \gls*{hpc} environment. In this case, a greedy algorithm is employed, at least in the theoretical development of the compiler (since to obtain the results, they applied a so-called ``naive'' algorithm, which is not specified). For the optimal qubit routing, they reach out for the work of t$|$ket$\rangle$ \cite{Sivarajah2021-tket}. Then, a distributed parallel execution is performed over the whole group of subcircuits employing the different devices, and, once the results are obtained, the CutQC work is again used to reconstruct the result of the original circuit using every result obtained in each subcircuit.

\paragraph{Compilers for embarrassing parallelism}
Despite the absence of compilers specifically designed for embarrassingly parallel tasks in quantum computing, the inherent parallelizable nature of these tasks --~primarily the distribution of shots across multiple \glspl*{qpu}~-- means that any quantum compiler or framework could be easily modified to support this mode of distribution. This adaptability is due to the fact that the distribution of computational tasks among different processors is a well-established practice in the field of \gls*{hpc}. Consequently, leveraging existing classical job distribution techniques allows for the straightforward parallel execution of quantum computations on multiple \glspl*{qpu}, highlighting a seamless integration of classical parallelism principles within quantum computing frameworks.

Nevertheless, an appreciation of the multi-programming case has to be made. Even though the already presented QuCloud and QuCloud+ \cite{liu2021qucloud, liu2024qucloud+} are considered mapping mechanisms, they possess a compilation task scheduler and could be naturally extended to be able to perform as compilers with a multi-programming approach. This is precisely the scope of \textit{palloq} system presented by Ohkura et al.\ \cite{ohkura2022simultaneous}, which includes a layout synthesis for multiple quantum circuits and a job scheduler to manage efficient and high fidelity quantum multi-programming. This compiler takes multiple quantum circuits written in OpenQASM \cite{Cross2017-OpenQASM} and the local gate error information of the device as input. Their layout synthesis employs a heuristic based on noise-adaptive layout, where the device's calibration data is analyzed to search for improved allocation using a greedy approach. Additionally, they propose a software-based crosstalk detection protocol utilizing a novel combination of randomized benchmarking methods to characterize the hardware's suitability for multi-programming.

\paragraph{Compilers combining types of distributions}
At the end of section \ref{compiler:subsec:types-of-distribution}, we mentioned the existence of a compiler that combines aspects of circuit distribution with the circuit-cutting technique \cite{Tomesh2021DivideComputation}. This work by Tomesh et al., as was already mentioned, introduced an algorithm called \gls*{qdca}. Among the main contributions of this work, there is the \gls*{qdca} specification, which contains several key elements: the partition of the input combinatorial optimization problem into multiple subproblems, the construction of the variational quantum circuit and the execution of it on distributed quantum computers using quantum circuit cutting techniques. The partition of the input is where the classical techniques of graph partitioning employed for circuit distribution take place, in this case, \gls*{kl} and METIS. Even though it is not circuit distribution \textit{per se}, it employs the graph partitioning techniques used in this kind of distribution to perform circuit cutting, which narrows the boundaries between these two approaches. This work presents quantum circuit cutting as a compilation tool within a hybrid, variational application. With this approach, they claimed to achieve approximate solutions to \gls*{mis} problems\footnote{The \gls*{mis} problem is a classic NP-Complete combinatorial optimization challenge defined on a graph $G=(V,E)$. Its objective is to identify the largest feasible independent set within $G$, where an independent set, denoted as $S \subset V$, consists of nodes that are not adjacent to each other.}.

%% file: Section5.tex
\section{Application layer}
\label{sec:Aplications}

In section \ref{compiler:subsec:types-of-distribution}, three different categories of quantum distribution were introduced based on the communication mechanisms available in the network: circuit distribution, circuit cutting, and embarrassingly parallel. This section describes some selected examples of applications using each execution mode.

\subsection{Circuit-distribution based applications}
\label{applications:subsection:circuit-distribution}

As mentioned in the introduction of the paper, one of the first distributed algorithms was proposed by Grover \cite{grover1997quantum}. In this work, he uses the circuit distribution with quantum communications to estimate the mean of $N$ numbers between -1 and 1 under ideal conditions. Later, Gupta et al.\  \cite{Gupta2007} present a distributed version of the Grover search algorithm using quantum communications. Initially, the algorithm is shown using only two \glspl*{qpu}, where an additional qubit is needed in each \gls*{qpu} to handle the quantum communications using an EPR pair. The complexity analysis shows that the classical Grover requirements for operations are maintained in this distributed version, since the increase in the number of operations due to the distribution scales with the number of qubits as in the original algorithm, but the number of classical communications per iteration is not increased. The paper does not show if the algorithm can scale to more than two \glspl*{qpu}. Cirac et al.\ \cite{Cirac1999} describe a distributed quantum phase estimation algorithm. 

One of the key quantum algorithms that present an exponential scaling is the Shor algorithm. The main drawback of this algorithm is the high number of qubits that are needed for a correct execution. Due to this requirement, it is a perfect candidate to use the circuit distribution technique. In \cite{Yimsiriwattana2004Shor}, a first proposal to use several \gls*{qpu} is made. Firstly, they show that the \gls*{qft} can be executed in parallel, substituting each controlled operation with a remote-controlled one. They also show that the modular exponentiation can be parallelized using a set of \glspl*{qpu}. Although a communication complexity of $\mathbb{O}((log_2N)^2)$ is needed, being $N$ the number of bits of the number to factorize, and the total number of qubits is increased, the size of each \gls*{qpu} is drastically reduced. 

More recently, Gidney et al.\ \cite{Gidney2021} analyze the hardware resources for factoring large numbers, using the Eker{\aa} and H{\aa}stad algorithm \cite{ekeraa2017quantum} instead of the Shor one. Applying several optimizations and taking into account the current methods for making logical qubits, they assert that a number of 2048 bits can be factorized in 8 hours with 20 million noisy qubits (if the operations work in the range of nanoseconds). However, due to the capabilities of the implemented additions needed to factorize the number, the qubits can be reduced to 11 million for each \gls*{qpu} when 2 are used and to 4 million for 8 \glspl*{qpu}. They require a quantum network with a low (but efficient) bandwidth of 150 qb/s. Later, Xiao et al.\ \cite{Xiao2023} present a parallel algorithm that reduce the number of needed qubits, dividing the algorithm between several \glspl*{qpu}, each one calculating one subset of the bits. Although the algorithm uses several \glspl*{qpu}, it is sequential because to guarantee that the correct state is used on each step, it is teletransported between them at the end of each step. 

More well-known quantum algorithms have been paralleli\-zed. For example, Neumann et al.\ \cite{Neumann2020ImperfectEstimation} study the Quantum Phase Estimation algorithm using a remote-controlled operation. They compared two possible approaches. The first one is called standard (or automatic), where each controlled operation in the standard \gls*{qft} is replaced by a remote-controlled operation. This case needs $n^2$ entangled pairs to execute. The second approach uses the iterative nature of the \gls*{qft}, aggregating all controlled operations by a single qubit in a unique transport operation. In this case, the number of transport operations is reduced to $n$. For the experiments, they used a simulator, introducing different noise levels in the creation of entangled pairs. The results obtained are similar for both approaches, given the last systematically better results. This experiment shows that automatic partitioning of the problems must take care of possible optimizations and multiple usage of a single pair. One important point is that they studied only the effect of imperfect entangling in the needed pairs, without taking into account other errors such as the measurement, controlled operations between the pairs and the \gls*{qpu} qubits, etc. 

Also, Van Meter et al.\ \cite{VanMeter2008ArithmeticMulticomputer} studied some of the possible arithmetic operations using teledata and telegate methods in different distributed topologies. They found that for these problems the teledata outperforms the telegate method and that a linear architecture is the best choice. 
In \cite{Tan2022DistributedProblem}, Tan et al.\ describe a parallel algorithm for Simon's problem that still keeps the exponential scaling when compared with the classical algorithm.

Recently, Li et al.\ \cite{Li2023DistributedProblem} present a family of distributed quantum algorithms for the classical Deutsch-Jozsa problem. These algorithms are based on a set of computers with remote communications. However, in the current description, the nature is still sequential, without a clear path to reduce the global depth and time. Finally, Shi et al. \cite{Shi2023} made a first proof of concept of using \gls*{qmpi} for the Quantum Phase Estimation and Trotter time evolution, but without including real quantum communications.

\subsection{Circuit knitting}
\label{applications:subsection:circuit-cutting}

As described in Sec.\ \ref{compiler:subsubsec:circuit-cutting}, algorithms based on circuit-cutting only need classical communications to calculate the final solution. Automatic cutting of a circuit (in space or time) is feasible when the number of control operations to cut is limited. 
However, it is also possible to find algorithms that divide a single problem (usually executed using a single quantum circuit) in the execution of several independent quantum programs that later must be combined classically to find the right solution, but using non-automatic clever designs. The set of techniques that allows dividing a quantum problem into subproblems, combining their independent results using classical post-processing to obtain the final result is called \emph{circuit knitting}.

As already mentioned in the introduction, the paper from Yepez \cite{Yepez2001type} was one of the first proposals to analyze this parallel computation in a hybrid scheme.
He considered the case of a system composed of quantum nodes but exclusively connected by a classical network. He named this architecture type-II quantum architecture to differentiate it from the monolithic quantum processors (of type-I), which maintain the global phase coherence.
The idea behind his proposal is that some problems need only short spatial and time entanglement, as some kinds of molecules. So they are tractable in parallel quantum computers, unlike other algorithms that need long and spatially large entanglement. For solving those problems, there are three assumptions: first, that the wave function is separable, i.e., can be expressed as a tensor product of subwave functions, each of them residing in one \gls*{qpu}; second, that we can apply a projection operator simultaneously on each qubit of each \gls*{qpu}; and, third, that this projection can be applied after each time step. Yepez proposes a quantum computer composed of many small QPUs arranged in a regular periodic lattice, where local operations are applied to the local qubits simultaneously across the lattice. He applies this proposal to solve problems with lattice gases. For small \glspl*{qpu}, maybe the problems could be tractable using modern Tensor Networks techniques.

In \cite{Zhou2023DistributedAlgorithm} and \cite{Zhou2023DistributedAlgorithmb}, Zhou et al.\ present distributed quantum algorithms for the Bernstein-Vazirani classical problem and the Grover search, respectively. They divide the binary functions used on the algorithms into a set of subfunctions that can be executed in parallel, getting the final result composing the different binary parts. In the case of Grover's search, the algorithm only works when a single solution exists, being still open the extension to multiple solutions. 
Similarly, Avron et al.\ \cite{Avron2021} study Deutsch-Jozsa's, Simon's, and Grover's on a distributed environment, finding that, for these algorithms, there are still advantages when comparing with the classical solutions, being the advantage reduced when compared with the fault-tolerant versions. But since these distributed algorithms require shallow circuits, they may be a short-term solution in today's \gls*{nisq} era.

The Iterative Quantum Phase Estimation, when the initial state is an eigenvector of the operator or Hamiltonian, can be executed in parallel: the different control operations of the powers of the unitary can be executed concurrently, needing only to communicate the final measure of the auxiliary qubit to the rest of the QPUs and combine the results to obtain the corresponding eigenvalue.

Several parallel versions of \glspl*{vqa} also use circuit-cutting techniques. For example, \cite{peng2020simulating} uses a circuit-cutting based \gls*{vqe} to calculate the ground state of BeH$_2$. Eddins et al.~\cite{Eddins2022forging} present another kind of methodology. They use the Schmidt decomposition to divide a chemical problem of $2N$ qubits in several circuits that need only $N$ qubits, applying \gls*{vqe} to those, and joining the results to calculate the final value of the observable. Fujii et al. \cite{Fujii2022} propose another method to divide the problem into smaller cases that are combined hierarchically to find the final solution. The technique can be applied when the problem has some structure that aggregates the entanglement in clusters that can be linked later at a higher level. They apply the technique to a kagome lattice, using several layers of aggregation. This technique could also be used in a hybrid scheme, where part of the calculation is done by \glspl*{qpu} at the first steps, and later, the system is solved by a classical computer using tensor networks.

The usage of these divide-and-conquer techniques can also be applied to combinatorial optimization, where a larger problem can be solved using several computers \cite{Tomesh2021DivideComputation,Zhou2022qaoa2}. 
The circuit cutting has also been applied to \gls*{qml}. Marshall et al.\ \cite{Marshall2022} examine it for the case of classification. They found that automatic circuit cutting could avoid executing all the subcircuits because some of them do not contribute significantly to the final result and propose a small change in the process that permits the achievement of results close to the classical Neural Networks in classification problems.

\subsection{Embarrassingly parallel applications}
\label{applications:subsection:Embarrasengly-parallel}
The cutting techniques presented in the previous section convert a complex problem into an example of an embarrassingly parallel application, where each smaller circuit can be executed in parallel, combining later the results classically. Other examples of these kinds of applications are \cite{Zhang2021-2, park2023quantum}, which study the use of partial diffusion operator \cite{Grover2002} for Grover's search algorithm. The use of this technique does not reduce the number of required qubits but presents some advantages because each circuit is smaller in depth (and consequently, needs less time to execute in parallel), and the angles of rotations are bigger, reducing the errors in current quantum devices.

Other quantum algorithms, such as the Phase Estimation for a single phase, can be executed using this formalism~\cite{Li2017ApplicationEstimation} because it is possible to split the algorithm into several smaller circuits and combine the results classically at the end. Other classical quantum algorithms, such as the Amplitude Estimation, require large resources that can be approximated by distributing several smaller tasks and post-processing classically their results~\cite{Tanaka2021}.

\subsection{Combined techniques}
\label{applications:subsection:Combined-techniques}

In order to get the maximum profit from the available distributed infrastructure or, in the short term, to permit the calculation of \glspl*{vqa}, a combination of the aforementioned techniques can be applied. For example, DiAdamo et al.\ \cite{net-DiAdamo2021} propose to place some of the needed circuits to calculate the expectation value on the available \glspl*{qpu} and use the remaining free qubits on them to make a distributed version defined of the Ansatz. Instead of using the circuit distribution version, another possibility could be splitting the Ansatz using the circuit cutting technique.

%% file: Conclusions.tex
\section{Conclusions}
\label{sec:conclusions}

Distributed quantum computing emerges as a clear pathway to enhance the computational capabilities of current quantum systems. In this work, we have presented a comprehensive survey of this field's current state of the art. Using a four-layered model --~physical, network, development, and application~--, we have guided readers to explore its foundational principles, achievements, challenges, and promising directions for further research. 

As it was explained, the most basic mechanism in the physical layer required for distributed algorithms in \gls*{dqc} applications is quantum teleportation. This resource enables the transmission of quantum states between qubits, regardless of their physical separation, thereby facilitating the creation of interconnected quantum processors. Two types of teleportation protocols can be defined: gate teleportation or telegate and qubit state teleportation or teledata. While the former enables the remote execution of quantum gates on entangled qubits, enabling the manipulation of quantum information without direct physical interaction, the second allows the unknown quantum state, processed in one network node, to be sent to a remote location. Enhancing the fidelity of these protocols is an active area of research, as it is crucial for ensuring quantum-computational accuracy in a future distributed quantum computer.

On a pure distributed architecture, where qubits are transported between \glspl*{qpu} or remote operations are employed, there are some initial results showing that the teledata could outperform the telegate method. Because this advantage could depend on the problem and on the techniques to make the teleoperation, more research is needed to confirm them. Also, because teledata could be executed using a single qubit for the transportation (instead of an EPR pair as was employed usually), this advantage could be exacerbated and simplify the final quantum network architecture.

However, to achieve truly interconnected, datacenter-scale \glspl*{qpu}, quantum networks must first be established in such a way that entanglement distribution is facilitated between any two nodes of the network. Current scalable proposals for entanglement distribution networks suggest the need for quantum networking devices, repeaters, switches, and routers, where entangled qubits for communication can be pre-established by transduction to flying qubits and successive entanglement distribution towards the end nodes, where the computation takes place. Quantum network devices must then have a register of qubits and implement a limited quantum operation instruction set necessary to carry out the entanglement distillation, swap, and teleportation protocols, unlocking true deterministic \gls*{dqc} architectures.

From an applicability and marketability standpoint, current networking solutions are costly and lack the performance/fidelity and robustness needed for a practical scenario. Higher-level aspects are still in the early stages of research, such as networking protocols, connectivity architectures, as well as scalability and robustness of the proposed solutions. Auxiliary protocols for synchronization, resource management for entanglement distribution, network services definition, error correction, and qubit encodings are yet to be developed to achieve the capabilities required for fault-tolerant, highly available, and performant networks suitable for DQC.

In the current noisy and limited \glspl*{qpu} scenario, circuit cutting can become a useful tool for solving large problems with small quantum computers, distributing parts of the circuit between them without needing a fully realized quantum network. However, the cost associated with this technique scales exponentially with the amount of cut (or, simulated) entanglement between the parts. For general quantum circuits, entanglement may have a very complex structure that is unknown beforehand. Clustered circuits with limited connectivity between the clusters are most promising in finding utility with circuit cutting.  
Some improvements have been proposed, and it may be possible to avoid the execution of a large fraction of the subcircuits, reducing the computing capability. However, there are some criticisms about the utility of these techniques.  
But dividing the circuits and executing them in different \glspl*{qpu} requires a better understanding of the effect of different noise profiles for each \gls*{qpu} and, when different architectures are employed, manage correctly the different times for execution. 

Using agnostic compilers to find the best partitions for a general algorithm is equivalent to the already-known concept of auto parallelism in classical computing, which is known to scale poorly. It can be better to design or choose problems that are easy to cut, such as well-designed ansatzes for variational quantum algorithms or problems adapted for modular architectures. Apart from the automatic tools for breaking the circuits, as in classical computing, wise programmers can find methods of dividing and parallelizing the algorithms. Tools for helping them to make implementations are needed, such as \gls*{qmpi} or frameworks that distribute the programs.

\section*{Acknowledgments} 

This work was supported by MICINN through the European Union NextGenerationEU recovery plan
(PRTR-C17.I1), and by the Galician Regional Government through the “Planes Complementarios de
I+D+I con las Comunidades Autónomas” in Quantum Communication. This work was also supported by the Ministry of Economy and Competitiveness, Government of Spain (Grant Numbers PID2019-104834GB-I00, PID2022-141623NB-I00 and PID2022-137061OB-
C22), Consellería de Cultura, Educación e Ordenación Universitaria (accreditations ED431C 2022/16 and ED431G-2019/04), and the European Regional Development Fund (ERDF), which acknowledges the CiTIUS-Research Center in Intelligent Technologies of the University of Santiago de Compostela as a Research Center of the Galician University System.